\documentclass[12pt]{article}
\usepackage{amsfonts,graphicx,amssymb,amsmath,amsthm,epsfig}
\usepackage{float}
\allowdisplaybreaks

\begin{document}
\title{\bf Compact Objects by Extended Gravitational Decoupling in $f(G,T)$ Gravity}
\author{M. Sharif$^1$ \thanks{msharif.math@pu.edu.pk} and K. Hassan$^2$
\thanks{komalhassan3@gmail.com}\\
$^1$ Department of Mathematics and Statistics, The University of Lahore,\\
1-KM Defence Road Lahore, Pakistan.\\
$^2$ Department of Mathematics, University of the Punjab,\\
Quaid-e-Azam Campus, Lahore-54590, Pakistan.}
\date{}

\maketitle

\begin{abstract}
In this paper, we investigate the anisotropic interior spherically
symmetric solutions by utilizing the extended gravitational
decoupling method in the background of $f(G,T)$ gravity, where $G$
and $T$ signify the Gauss-Bonnet term and trace of the stress-energy
tensor, respectively. The anisotropy in the interior geometry arises
with the inclusion of an additional source in the isotropic
configuration. In this technique, the temporal and radial potentials
are decoupled which split the field equations into two independent
sets. Both sets individually represent the isotropic and anisotropic
configurations, respectively. The solution corresponding to the
first set is determined by using the Krori-Barua metric potentials
whereas the second set contains unknown which are solved with the
help of some constraints. The ultimate anisotropic results are
evaluated by combining the solutions of both distributions. The
influence of decoupling parameter is examined on the matter
variables as well as anisotropic factor. We illustrate the viable
and stable features of the constructed solutions by using energy
constraints and three stability criteria, respectively. Finally, we
conclude that the obtained solutions are viable as well as stable
for the whole domain of the coupling parameter.
\end{abstract}

\section{Introduction}

The gigantic cosmos contains systematic structures ranging from
small bodies to massive configurations like clouds, stars, clusters,
super-clusters and galaxies. A widely recognized theory, general
relativity (GR), has played an essential role in comprehending the
mysterious features and evolution of the universe. It is assumed
that our cosmos is comprised of ordinary source, dark matter and
dark energy. The visible part of the universe is ordinary matter,
while dark matter and dark energy have some ambiguous and enigmatic
nature, which are supposed to be well delineated by GR. Further, it
helps in resolving the velocity curves of galaxies \cite{1} together
with accelerated cosmic expansion \cite{2}. The presence of dark
energy was explicated by accommodating the cosmological constant
into the Lambda cold dark matter ansatz. Nevertheless, the
readjustment of the values of cosmological constant is highly needed
in order to describe the dynamics of the universe through several
cosmic eras and its matching with the observational data. Thus, to
resolve these issues, modified gravity theories are regarded as the
favorable alternatives to GR. The Einstein-Hilbert action is altered
to obtain the modified theories by either adding or replacing the
scalar curvatures and their related generic functions.

The forthright generalization of GR in higher dimensions is the
Lovelock theory of gravity, which becomes equivalent to GR in
4-dimensions \cite{2a}. This theory yields two scalars, the first
one is the Ricci scalar $R$ (also called as the first Lovelock
scalar), and the second one is the Guass-Bonnet invariant (GB)
(dubbed as the second Lovelock scalar). Another way of modifying
action is achieved with the help of second lovelock scalar (GB
invariant) which gives rise to Einstein GB gravity in 5-dimensions
\cite{2b}. The GB term in mathematical notation is denoted as
\begin{equation}\nonumber
G=R^{\varsigma\sigma\nu\mu}R_{\varsigma\sigma\nu\mu}
-4R^{\varsigma\sigma}R_{\varsigma\sigma}+R^2,
\end{equation}
which is presented as a conjunction of the curvature scalar, Ricci
tensor $(R_{\varsigma\sigma})$ and curvature tensor
$(R_{\varsigma\sigma\nu\mu})$. It is a 4-dimensional invariant and
free from spin-2 ghost instabilities. To understand the effects of
GB invariant in four dimensions, Nojiri and Odintsov \cite{3}
modified the Einstein-Hilbert action by including the generic
function $f(G)$ which led to $f(G)$ gravity or modified GB theory.
This gravity is supposed to investigate the progression from
decelerated to accelerated phase as well as adequately describes the
salient aspects of cosmic expansion.

One of the simplest extension of GR is introduced by substituting
$R$ with its generic $f(R)$ function in the Lagrangian, namely
$f(R)$ gravity. Several researchers utilized the feasible $f(R)$
models to examine the inflationary and cosmic acceleration of the
universe \cite{32}. Bertolami et al \cite{31} first proposed the
concept of matter-geometry coupling in $f(R)$ theory, using
Lagrangian as a function of $R$ and $\pounds_{m}$ to investigate the
effects of this interaction on massive objects. This interaction
prompted many researchers to focus their attention on proposing a
coupling that helps in studying the fast cosmic expansion
efficiently. In this respect, Harko et al \cite{33} coupled matter
and geometric expressions in the Einstein-Hilbert action and
introduced $f(R,T)$ theory.

Sharif and Ikram \cite{8} proposed another non-minimal coupled
gravity, i.e., $f(G,T)$ theory and discussed energy conditions in
FRW universe. In this theory, the energy-momentum tensor (EMT) is
not conserved and test particles follow the non-geodesic track as a
result of an extra force. The addition of $T$ along with $G$
significantly demonstrates the fascinating outcomes regarding the
present cosmos. The same authors \cite{8a} studied the stability of
some cosmological models via linear perturbation in the realm of
isotropic and homogeneous cosmos. Mustafa et al. \cite{8e} examined
the necessary physical properties of three compact objects
possessing anisotropic configurations and obtained well-behaved
solutions in $f(G,T)$ gravity. We have decomposed the Riemann tensor
using Herrera approach to evaluate the complexity factor in the
static cylindrical structure (uncharged-charged), which was further
discussed for non-static uncharged and charged spherical as well as
cylindrical geometries \cite{8d}.

In dense compact entities, the interactions of substances exhibit
distinct characteristics in different directions that ensure the
existence of anisotropy within the compact structures \cite{9}.
Anisotropy in the inner configurations is believed to be induced by
phase transition \cite{9a} and superfluid \cite{9b}. Herrera and
Santos \cite{9c} looked at the causes of anisotropy as well as how
it affected the progression of astrophysical objects. By using a
specific anisotropic factor, Harko and Mak \cite{10} were able to
study the anisotropic static spherical structures through the
analytical solution of the field equations. Dev and Gleiser
\cite{10a} determined exact solutions of the field equations using
different forms of the equation of state relating tangential as well
as radial pressure and examined the remarkable influence of pressure
anisotropy on physical attributes of celestial objects. Paul and Deb
\cite{10b} studied the anisotropic stellar entities in hydrostatic
equilibrium. Arbanil and Malheiro \cite{10c} investigated the
stability of anisotropic strange stars through numerical solutions
by employing MIT bag model.

There exist a number of past related works on the solutions of the
gravitational field equations in different modified theories which
can be used to model physically acceptable compact bodies. In GR,
Errehymy et al \cite{34} studied the substantial features of
anisotropic celestial bodies which were found to be less dense.
Moreover, it can be observed that along with the less dense stellar
stars in GR, only the radial component of adiabatic index is
utilized in evaluating their realistic configurations \cite{35}. The
graphical analysis of compact stars in $f(G)$ gravity is assessed
without utilizing the adiabatic index criterion by several
researchers \cite{36}. A similar pattern is followed by Shamir and
Zia \cite{37} in the framework of $f(R,G)$ gravity.

The vague nature of astrophysical systems is obtained through
analytic solutions of the field equations. The field equations
contain several geometric ingredients and are highly nonlinear,
making it challenging to compute such solutions. Due to the
non-linear behavior, researchers have always been interested to
develop specific procedure that can be utilized to solve these
equations and provide physically feasible results. In order to
address this issue, a recently developed technique so called
gravitational decoupling via minimal geometric deformation (MGD) has
proven helpful in determining the feasible anisotropic solutions.
Primarily, Ovalle \cite{11} implemented this technique to find new
anisotropic spherical solutions. Afterwards, Ovalle et al. \cite{13}
calculated the anisotropic domains by generalizing the isotropic
system and analyzed them graphically. Gabbanelli et al. \cite{14}
worked on Durgapal-Fuloria solution to compute the anisotropic
solution.

Estrada and Tello-Ortiz \cite{17} employed the gravitational
decoupling to formulate new analytic anisotropic stellar models and
graphically analyze their physical features. Singh et al. \cite{18}
investigated interior anisotropic solutions in class-I spacetime and
estimated the radius along with mass of compact bodies through
$M$-$R$ curve. Hensh and Stuchlik \cite{19} utilized the Tolman VII
solution as a seed source for computing its anisotropic version.
Sharif and Saba \cite{16} used the known solution for the perfect
source and explored the physical properties of the charged-uncharged
anisotropic domains in $f(G)$ gravity. A substantial body of
research has been done to produce the anisotropic interior solutions
using this technique in different modified theories \cite{21}.
Recently, we have discussed new anisotropic models corresponding to
Tolman V, Krori-Barua ansatz and Karmarkar condition for uncharged
and charged spherical geometries through MGD. Further, the extended
geometric deformation method has also been applied to analyze the
constructed anisotropic solutions in $f(G,T)$ theory \cite{21a}.

In MGD technique, only the radial coefficient is distorted while
keeping the temporal part unperturbed, hence leads to some
limitations. In this approach, there is no energy transmission
between matter sources, so the only interaction is gravitational.
Ovalle \cite{21b} proposed an extension of MGD which also decouples
the temporal coordinate along with the radial, termed as extended
gravitational decoupling (EGD). Contreras and Bargueno \cite{21c}
applied this approach for 2+1 dimensional spacetime to extend the
charged BTZ solution by addressing vacuum BTZ solution. Sharif and
Ama-Tul-Mughani \cite{21d} deformed both the metric functions to
construct two anisotropic solutions from a known Tolman IV and
Krori-Barua perfect fluid source. Similarly, Sharif and Saba
\cite{21e} studied salient features of the resulted anisotropic
solutions in $f(G)$ gravity using Tolman IV as seed sector.

\section{Essence of $f(G,T)$ Theory}

In $f(G,T)$ gravity, the action integral to formulate the field
equations is given as
\begin{equation}\label{1}
\mathbb{S}_{f(G,T)}=\int\bigg[\frac{\mathrm{R}+f(G,T)}{16\pi}
+\pounds_{m}+\alpha\pounds_{\delta}\bigg]\sqrt{-g}d^{4}x,
\end{equation}
where $g$ denotes the determinant of the metric tensor and
$\pounds_{m}$ stands for the matter Lagrangian density. Here, the
matter lagrangian density is taken as the positive pressure
\cite{24a} and $\pounds_{\delta}$ denotes the Lagrangian density
corresponding to the extra sector. The relationships defining the
lagrangian densities with their EMT sources are as follows
\begin{align}\label{1a}
T_{\varsigma\sigma}=g_{\varsigma\sigma}\pounds_m-\frac{2\partial\pounds_m}{\partial
g^{\varsigma\sigma}}, \quad
\delta_{\varsigma\sigma}=g_{\varsigma\sigma}\pounds_{\delta}-\frac{2\partial\pounds_{\delta}}{\partial
g^{\varsigma\sigma}}.
\end{align}
Here, the action (1) is varied with respect to the metric tensor to
develop the field equations corresponding to $f(G,T)$ gravity in the
following form
\begin{equation}\label{3}
G_{\varsigma\sigma}=8\pi
T^{\textsf{(tot})}_{\varsigma\sigma}=8\pi(T^{\textsf{(Cor)}}_{\varsigma\sigma}
+T^{\textsf{(M)}}_{\varsigma\sigma}+\alpha\delta_{\varsigma\sigma}),
\end{equation}
where $\alpha$ expresses the decoupling parameter and
$G_{\varsigma\sigma}=R_{\varsigma\sigma}-\frac{1}{2}R
g_{\varsigma\sigma}$ represents the Einstein tensor. The extra
gravitational source $\delta_{\varsigma\sigma}$ induces anisotropy
in the current configuration, and decoupling parameter $\alpha$
connects the seed and additional sectors. Moreover, the extra
curvature terms of $f(G,T)$ theory read
\begin{eqnarray}\nonumber
T^{\textsf{(Cor)}}_{\varsigma\sigma}&=&\frac{1}{8\pi}
\bigg[\{(p+\rho)\upsilon_{\varsigma}\upsilon_{\sigma}\}f_{T}(G,T)+\frac{g_{\varsigma\sigma}f(G,T)}{2}
+\big(4R^{\mu\nu}R_{\varsigma\mu \sigma \nu}\\\nonumber&-&
2RR_{\varsigma\sigma}-2R^{\mu \nu \gamma} _{\varsigma}R_{\sigma \mu
\nu \gamma}+4R_{\mu\sigma}R^{\mu}_{\varsigma}\big)f_{G}(G,T)
+(4g_{\varsigma\sigma}R^{\mu
\nu}\nabla_{\mu}\nabla_{\nu}\\\nonumber&-&4R^{\mu}_{\varsigma}\nabla_{\sigma}\nabla_{\mu}-4R_{\varsigma\mu
\sigma\nu}\nabla^{\mu}\nabla^{\nu}-2g_{\varsigma\sigma}R\nabla^{2}
+2R\nabla_{\varsigma}\nabla_{\sigma}-4R^{\mu}_{\sigma}\nabla_{\varsigma}\nabla_{\mu}\\\label{4}&+&4R_{\varsigma\sigma}\nabla^{2})f_{G}(G,T)\bigg],
\end{eqnarray}
where the d' Alembert operator is indicated by
$\Box=\nabla^{a}\nabla_{a}=\nabla^{2}$ and
$\Theta_{\varsigma\sigma}=-2T_{\varsigma\sigma}+pg_{\varsigma\sigma}$.
The partial derivatives of $f(G,T)$ with respect to $G$ and $T$ are
denoted by $f_G=\frac{\partial f(G,T)}{\partial G}$ and
$f_T=\frac{\partial f(G,T)}{\partial T}$, respectively. A
consequential role is played by EMT to disclose the interior
configuration of the self-gravitating entities. The EMT for the
perfect matter source filled in the internal regime is described by
\begin{equation}\label{5}
T^{\textsf{(M)}}_{\varsigma\sigma} =(\rho+p)
\upsilon_{\varsigma}\upsilon_{\sigma}+pg_{\varsigma\sigma},
\end{equation}
where $\upsilon_{\varsigma}$ depicts the four-velocity satisfying
the relation $\upsilon^{\varsigma}\upsilon_{\varsigma}=-1$, $p$ and
$\rho$ demonstrate the pressure and density, respectively.

The geometry under consideration is composed of inner and outer
regions divided by the hypersurface. The internal spherically
symmetric structure (static) is defined by the following metric
\begin{equation}\label{6}
ds^{2}=-e^{\varphi}dt^{2}+e^{\vartheta}dr^{2}+r^{2}(d\theta^{2}+{\sin^{2}\theta}{d\phi^2}),
\end{equation}
where both $\varphi$ and $\vartheta$ are functions of $r$ solely.
The velocity in terms of its components is written as
\begin{equation}\label{6a}
\upsilon^{\varsigma}=\left(e^{\frac{-\varphi}{2}},0,0,0\right).
\end{equation}
For self-gravitating astrophysical objects, the modified field
equations are
\begin{eqnarray}\label{8}
8\pi(\widetilde{\rho}+T^{0\textsf{(Cor)}}_{0}-\alpha\delta^{0}_{0})
&=&
\frac{1}{r^{2}}+e^{-\vartheta}(\frac{\vartheta'}{r}-\frac{1}{r^{2}})
,\\\label{9}
8\pi(\widetilde{p}+T^{1\textsf{(Cor)}}_{1}+\alpha\delta^{1}_{1}) &=&
-\frac{1}{r^{2}}+e^{-\vartheta}(\frac{1}{r^{2}}+\frac{\varphi'}{r}),
\\\label{10}
8\pi(\widetilde{p}+T^{2\textsf{(Cor)}}_{2}+\alpha\delta^{2}_{2})
&=&e^{-\vartheta}(\frac{\varphi'^{2}}{4}+\frac{\varphi''}{2}-\frac{\vartheta'\varphi'}{4}
-\frac{\vartheta'}{2r}+\frac{\varphi'}{2r}),
\end{eqnarray}
where the derivative with respect to $r$ is specified by prime and
\begin{align}\label{10a}
\widetilde{p}=p+\frac{\psi}{16\pi}(-\rho+3p),\quad
\widetilde{\rho}=\rho+\frac{\psi}{16\pi}(3\rho-p),
\end{align}
correction terms $T^{0\textsf{(Cor)}}_{0},~T^{1\textsf{(Cor)}}_{1}$
and $T^{2\textsf{(Cor)}}_{2}$ are exhibited in Appendix \textbf{A}
(Eqs.\eqref{61}-\eqref{63}).

The extra force exists because the EMT is not conserved in this
theory. Meanwhile, the non-conservation of matter configuration is
characterized by the following equation
\begin{eqnarray}\nonumber
\nabla^{\varsigma}T_{\varsigma\sigma}&=&\frac{f_{T}(G,T)}{8\pi-f_{T}(G,T)}
\bigg[\nabla^{\varsigma}\Theta_{\varsigma\sigma}-\frac{1}{2}g_{\varsigma\sigma}\nabla^{\varsigma}T
+(\Theta_{\varsigma\sigma}+T_{\varsigma\sigma})\nabla^{\varsigma}(\ln
f_{T}(G,T))\bigg],
\end{eqnarray}
which is left with the non-zero term
\begin{align}\label{12}
\frac{dp}{dr}+\frac{\vartheta'}{2}(p+\rho)+\alpha\frac{d\delta^{1}_{1}}{dr}
+\frac{2\alpha}{r}(\delta^{1}_{1}-\delta^{2}_{2})+\frac{\alpha\vartheta'}{2}(\delta^{1}_{1}-\delta^{0}_{0})=\Gamma,
\end{align}
where $\Gamma$ includes the modified terms as
\begin{equation}\label{66}
\Gamma=\frac{\psi}{8\pi-\psi}\bigg[-\frac{(-\rho+3p)'}{2}-\alpha\delta^{1}_{1}(\ln
f_T)'+(-2p)'\bigg].
\end{equation}
Here, it is important to note that a successful decoupling is
achieved in EGD approach when the exchange of energy between normal
matter and extra source happens. The following explicit $f(G,T)$
theory model \cite{23a} is used to explore the viable and stable
anisotropic solutions
\begin{equation}\label{60}
f(G,T)= \mathfrak{f_1}(G)+\mathfrak{f_2}(T),
\end{equation}
where $\mathfrak{f_1}$ and $\mathfrak{f_2}$ are separately defined
functions of $G$ and $T$, respectively. In this curvature-matter
coupled theory, we select a quadratic model to analyze its role in
understanding the physical features of the stellar structure. Hence,
$\mathfrak{f_1}(G)=\chi G^2$ and $\mathfrak{f_2}(T)=\psi T$ are
fixed, where $\psi$ and $\chi$ are free parameter and real constant,
respectively. The expressions of $G$ along with its higher
derivatives are exhibited in Eqs.\eqref{63a}-\eqref{65} of Appendix
\textbf{A}.

The non-linear differential Eqs.\eqref{8}-\eqref{10} as well as
\eqref{12} form a system with seven unknown quantities
$(\varphi,\vartheta,\rho,p,\delta^{0}_{0},\delta^{1}_{1},\delta^{2}_{2})$,
indicating that the system has fewer equations than unknown
parameters. Thus, to close the system more constraints are required.
For this purpose, we use a systematic scheme of EGD to obtain the
solution of our system. The matter variables are easily identified
as
\begin{equation}\label{13}
\check{\rho}=\rho-\alpha\delta^{0}_{0},\quad
\check{p_r}=p+\alpha\delta^{1}_{1},\quad
\check{p_t}=p+\alpha\delta^{2}_{2}.
\end{equation}
The above expressions assure that the anisotropy is induced due to
the extra source ($\delta^{\varsigma}_{\sigma}$) within
self-gravitating system . Thus, when
$\delta^{1}_{1}\neq\delta^{2}_{2}$, the effective anisotropy becomes
\begin{equation}\label{14}
\check{\Delta}=\check{p_t}-\check{p_r}=\alpha(\delta^{2}_{2}-\delta^{1}_{1}).
\end{equation}

\section{Extended Gravitational Decoupling Scheme}

In this section, a novel approach entitled as gravitational
decoupling by means of EGD is utilized to determine the unknowns by
resolving the system \eqref{8}-\eqref{10}. According to this method,
the field equations are segregated such that the anisotropy produced
in the internal structure is caused by the presence of extra source
($\delta^{\varsigma}_{\sigma}$). For this purpose, we consider the
following metric for perfect matter source
\begin{equation}\label{15}
ds^{2}=\frac{dr^{2}}{\epsilon(r)}-e^{\xi(r)}dt^{2}+r^{2}d\theta^{2}+r^{2}{\sin^{2}\theta}{d\phi^2},
\end{equation}
where $\epsilon(r)=1-\frac{2m}{r}$, $m(r)$ conforms the Misner-Sharp
mass of the inner celestial structure. The effects of anisotropy on
the perfect source are encoded by implementing the linear
geometrical transformation to temporal as well as radial metric
components through
\begin{equation}\label{16}
\xi\rightarrow\varphi=\xi+\alpha h^\ast,\quad \epsilon\rightarrow
e^{-\vartheta(r)}=\epsilon+\alpha k^\ast,
\end{equation}
where $k^\ast$ and $h^\ast$ are the deformation functions associated
to radial and temporal metric potentials, respectively, and $\alpha$
participates in governing the working of both deformations. These
decompositions divide the field equations \eqref{8}-\eqref{10} into
two arrays, in which the first set represents the perfect source
$(\alpha=0)$ as
\begin{align}\label{18}
8\pi(\rho+\frac{\psi}{16\pi}(3\rho-p)+T^{0\textsf{(Cor)}}_{0})
&=\frac{1}{r^{2}}-(\frac{\epsilon'}{r}+\frac{\epsilon}{r^{2}}),
\\\label{19}
8\pi(p+\frac{\psi}{16\pi}(-\rho+3p)+T^{1\textsf{(Cor)}}_{1})
&=-\frac{1}{r^{2}}+\frac{\epsilon}{r}(\frac{1}{r}+\xi'),
\\\nonumber
8\pi(p+\frac{\psi}{16\pi}(-\rho+3p)+T^{2\textsf{(Cor)}}_{2})&=\epsilon(\frac{\xi''}{2}+\frac{\xi'^{2}}{4}+\frac{\xi'}{2r})\\\label{20}&+\epsilon'(\frac{\xi'}{4}
+\frac{1}{2r}).
\end{align}

Solving the above equations, we obtain density and pressure for the
isotropic sector as
\begin{align}\nonumber
\rho&=\frac{-1}{4 \left(\psi ^2+12 \pi \psi +32 \pi ^2\right)
r^2}\bigg(-2 \psi +3 \psi  r^2 T^{0\textsf{(Cor)}}_{0}+16 \pi r^2
T^{0\textsf{(Cor)}}_{0}\\\label{20a}&+\psi  r^2
T^{1\textsf{(Cor)}}_{1} +3 \psi r \epsilon '-\psi  r \epsilon \xi
'+2 \psi  \epsilon+16 \pi r \epsilon '+16 \pi  \epsilon-16 \pi
\bigg),
\\\nonumber
p&=\frac{-1}{4 \left(\psi ^2+12 \pi \psi +32 \pi ^2\right)
r^2}\bigg(2 \psi +\psi  r^2 T^{0\textsf{(Cor)}}_{0}+3 \psi  r^2
T^{1\textsf{(Cor)}}_{1}\\\label{20b}&+16 \pi  r^2
T^{1\textsf{(Cor)}}_{1} +\psi r \epsilon '-3 \psi  r \epsilon \xi
'-2 \psi  \epsilon -16 \pi r \epsilon \xi '-16 \pi  \epsilon+16 \pi
\bigg),
\end{align}
whereas the induced anisotropy due to new sector is evaluated
through the second set
\begin{eqnarray}\label{21}
8\pi\delta^{0}_{0}&=&\frac{k^{\ast'}}{r}+\frac{k^{\ast}}{r^2},
\\\label{22}
8\pi\delta^{1}_{1} &=&\frac{\alpha  k^\ast
h^{\ast'}}{r}+\frac{\epsilon
h^{\ast'}}{r}+\frac{k^\ast}{r^2}+\frac{k^\ast \xi '}{r},
\\\nonumber
8\pi\delta^{2}_{2}&=&\frac{k^{\ast'} \big(\alpha
h^{\ast'}+\xi'\big)}{4} +\frac{h^{\ast'} \epsilon '}{4}  +k^\ast
\bigg(\frac{\big(\alpha h^{\ast''}+\xi''\big)}{2} +\frac{\big(\alpha
h^{\ast'}+\xi'\big)^{2}}{4}
\\\label{23}&+&\frac{\alpha
h^{\ast'}+\xi'}{2 r}\bigg)+\frac{k^{\ast'}}{2 r}+\epsilon
\bigg(\frac{ h^{\ast''}}{2} +\frac{\alpha h^{\ast' 2}}{4} +\frac{
h^{\ast'} \xi'}{2} +\frac{ h^{\ast'}}{2 r}\bigg).
\end{eqnarray}
It should be noticed that the system \eqref{18}-\eqref{20} is
comprised of four unknowns, i.e., $p, \rho, \xi$ and $\epsilon$,
while the anisotropic set \eqref{21}-\eqref{23} contains nine
unknowns $(\rho, p, \xi,\epsilon,
\delta^{0}_{0},\delta^{1}_{1},\delta^{2}_{2},k^\ast,h^\ast)$. Thus
we need to specify the isotropic set so that we might be able to
determine the solution corresponding to anisotropic sector.
Consequently, the number of unknowns will be reduced from seven to
five, indicating that the EGD technique assists in developing the
anisotropic solutions.

\section{Interior Anisotropic Solutions}

Now, we study anisotropic solutions of the astrophysical object
through some definite forms of the isotropic set. For this purpose,
we take the Krori-Barua metric \cite{22} as seed (isotropic)
solution which is attributed to singularity-free nature. The metric
coefficients are defined as
\begin{equation}\label{32}
e^{\xi(r)}=e^{L r^{2}+P},\quad e^{\vartheta(r)}=\epsilon^{-1}=e^{X
r^{2}}.
\end{equation}
Consequently, the values of $\rho$ and $p$ from
Eqs.\eqref{18}-\eqref{20} in terms of the above mentioned potentials
become
\begin{align}\nonumber
\rho&=\frac{e^{-r^2 X}}{4 \big(\psi ^2+12 \pi  \psi +32 \pi ^2\big)
r^2}\bigg( \big(-\psi  \big(-2 L r^2+e^{r^2 X} \big(r^2 (3
T^{0\textsf{(Cor)}}_{0}+T^{1\textsf{(Cor)}}_{1})-2\big)\\\label{33a}&-6
r^2 X+2\big)-16 \pi \big(\big(r^2 T^{0\textsf{(Cor)}}_{0}-1\big)
e^{r^2 X}-2 r^2 X+1\big)\big)\bigg),
\\\nonumber
p&= \frac{e^{-r^2 X}}{4 \big(\psi ^2+12 \pi \psi +32 \pi ^2\big)
r^2}\bigg( \big(-\psi  \big(e^{r^2 X} \big(r^2
(T^{0\textsf{(Cor)}}_{0}+3 T^{1\textsf{(Cor)}}_{1})+2\big)-2 \big(3
L r^2\\\label{33b}&+r^2 X+1\big)\big)-16 \pi  \big(-2 L r^2+\big(r^2
T^{1\textsf{(Cor)}}_{1}+1\big) e^{r^2 X}-1\big)\big)\bigg).
\end{align}
The junction conditions help to determine the unknown constants
$(L,P,X)$ involved in the above equations. When the interior and
exterior structures are matched over the hypersurface, the
continuity of metric potentials give the constants as
\begin{align}\label{36}
L&=\frac{\mathbb{M}_o \textsf{R}}{\textsf{R}^4 \left(1-\frac{2
\mathbb{M}_o}{\textsf{R}}\right)},\\\label{36a}
P&=\frac{\textsf{R}^2 \left(1-\frac{2
\mathbb{M}_o}{\textsf{R}}\right) \ln \left(1-\frac{2
\mathbb{M}_o}{\textsf{R}}\right)-\mathbb{M}_o
\textsf{R}}{\textsf{R}^2 \left(1-\frac{2
\mathbb{M}_o}{\textsf{R}}\right)},\\\label{36b}
X&=\frac{1}{\textsf{R}^2}\ln\bigg(\frac{1}{1-\frac{2\mathbb{M}_o}{\textsf{R}}}\bigg).
\end{align}
together with  $\frac{\mathbb{M}_o}{\textsf{R}}<\frac{4}{9}$,
$\mathbb{M}_o$ denotes the mass at the boundary. The compatibility
of isotropic solution \eqref{33a} and \eqref{33b} with the
Schwarzschild at the junction is assured by the above-mentioned
equations that can be altered in the internal regime with the
addition of extra source. The radial and temporal coefficients
(Eq.\eqref{32}) will help to evaluate the anisotropic solutions. The
deformation functions $h^\ast$ and $k^\ast$ are related to the
source term in \eqref{21}-\eqref{23}. The solution of this system is
evaluated by implying certain constraints. As this system
constitutes five unknowns and three equations, therefore, we require
two more constraints to close the system. We utilize the linear
equation of state on $\delta^{\varsigma}_{\sigma}$ as
\begin{equation}\label{36c}
\delta^{0}_{0}=a\delta^{1}_{1}+b\delta^{2}_{2}.
\end{equation}
For the sake of simplicity, we substitute $a=1$ and $b=0$, hence,
the above equation will become $\delta^{0}_{0}=\delta^{1}_{1}$.
Furthermore, we use the astrophysical object 4U 1820-30 \cite{22a}
whose mass and radius are $1.58\pm0.06M_\odot$ and $9.1\pm0.4$km,
respectively.

In the subsequent sections, we implement some limitations to develop
two interior anisotropic solutions and will then analyze their
graphical behavior.

\subsection{The First Solution}

The system \eqref{21}-\eqref{23} is closed by imposing an additional
condition on the radial part of the new source together with the
equation of state. These two constraints are utilized to determine
the deformation functions $(h^\star,k^\star)$ which are further
employed in formulating the components of
$\delta^{\varsigma}_{\sigma}$. It can be noticed that the inner
configuration depicts the consistency with outer distribution as far
as
$\widetilde{p}(\textsf{R})+T^{1\textsf{(Cor)}}_{1}(\textsf{R})\sim\alpha(\delta^{1}_{1}(\textsf{R}))_{-}$
holds. This requirement is fulfilled by using \cite{13}
\begin{equation}\label{37}
\widetilde{p}+T^{1\textsf{(Cor)}}_{1}=\delta^{1}_{1}.
\end{equation}
By utilizing the metric coefficients \eqref{32} in the field
equations \eqref{19},\eqref{21} and \eqref{22}, the deformation
functions become
\begin{align}\nonumber
h^\ast&=\int\big\{\sqrt{\pi } \big(2 L r^2+1\big) (L+X) e^{r^2 X}
\text{Erf}\big(r \sqrt{X}\big)-2 r \sqrt{X} \big(2 L^2 r^2+L \big(2
r^2 X \\\nonumber&\big(e^{r^2 X}+1\big)+1\big)+X\big)\big\}\big\{r
\big(2 r \sqrt{X} \big(\alpha  L+X \big(\alpha  e^{r^2
X}-1\big)\big)-\sqrt{\pi } \alpha  (L+X) \\\label{38a}&e^{r^2 X}
\text{Erf}\big(r \sqrt{X}\big)\big)\big\}^{-1} \, dr,\\\label{38b}
k^\ast&=\frac{\sqrt{\pi } (L+X) \text{Erf}\big(r \sqrt{X}\big)}{2 r
X^{3/2}}-\frac{L e^{-r^2 X}+X}{X}.
\end{align}
These deformation functions help to compose the temporal and radial
potentials as
\begin{align}\nonumber
\varphi&=L r^2+P+\alpha \int\big\{\sqrt{\pi } \big(2 L r^2+1\big)
(L+X) e^{r^2 X} \text{Erf}\big(r \sqrt{X}\big)-2 r \sqrt{X} \big(2
L^2 r^2\\\nonumber&+L \big(2 r^2 X \big(e^{r^2
X}+1\big)+1\big)+X\big)\big\}\big\{r \big(2 r \sqrt{X} \big(\alpha
L+X \big(\alpha  e^{r^2 X}-1\big)\big)\\\label{38c}&-\sqrt{\pi }
\alpha (L+X) e^{r^2 X} \text{Erf}\big(r
\sqrt{X}\big)\big)\big\}^{-1} \, dr,\\\nonumber
e^{-\vartheta}&=-\alpha +\frac{\sqrt{\pi } \alpha (L+X)
\text{Erf}\big(r \sqrt{X}\big)}{2 r X^{3/2}}+\frac{e^{-r^2 X}
(X-\alpha  L)}{X},
\end{align}
and for $\alpha=0$, this yields the standard Krori-Barua solution
for perfect source.

We employ the junction conditions to examine the effect of
anisotropy on $L,P$ and $X$. Thus, the first fundamental form of
matching conditions yields the following results
\begin{align}\nonumber
\ln\bigg(1-\frac{ \mathbb{M}_o}{\textsf{R}}\bigg)&=L
\textsf{R}^2+P+\alpha \bigg[\int\big\{\sqrt{\pi } \big(2 L
r^2+1\big) (L+X) e^{r^2 X} \text{Erf}\big(r
\sqrt{X}\big)\\\nonumber&-2 r \sqrt{X} \big(2 L^2 r^2+L \big(2 r^2 X
\big(e^{r^2 X}+1\big)+1\big)+X\big)\big\}\big\{r \big(2 r
\sqrt{X}\\\label{39}& \big(\alpha L+X \big(\alpha  e^{r^2
X}-1\big)\big)-\sqrt{\pi } \alpha (L+X) e^{r^2 X} \text{Erf}\big(r
\sqrt{X}\big)\big)\big\}^{-1} \,
dr\bigg]_{r=\textsf{R}},\\\label{40}
1-\frac{\mathbb{M}_o}{\textsf{R}}&= \frac{\sqrt{\pi } \alpha (L+X)
\text{Erf}\big(\textsf{R} \sqrt{X}\big)}{2 \textsf{R}
X^{3/2}}+\frac{e^{-\textsf{R}^2 X} (X-\alpha  L)}{X}-\alpha.
\end{align}
In the same way, the second fundamental form
$\big(\widetilde{p}(\textsf{R})+T^{1\textsf{(Cor)}}_{1}(\textsf{R})-\alpha(\delta^{1}_{1}(\textsf{R}))_{-}=0\big)$
gives
\begin{equation}\label{41}
X=\frac{\ln(1+2L \textsf{R}^2)}{\textsf{R}^2}.
\end{equation}
The necessary and sufficient conditions to match the interior and
exterior geometries at the junction are provided by
Eqs.\eqref{39}-\eqref{41}. The expressions of first anisotropic
solution and anisotropic factor corresponding to \eqref{36c} and
\eqref{37} are
\begin{align}\nonumber 
\check{\rho}&=\frac{e^{-r^2 X}}{8 r^2}\bigg[\frac{1}{\psi ^2+12 \pi
\psi +32 \pi ^2}\big\{-2 \psi \big(-2 L r^2+e^{r^2 X} \big(r^2 (3
T^{0\textsf{(Cor)}}_{0}+T^{1\textsf{(Cor)}}_{1})\\\nonumber&-2\big)-6
r^2 X+2\big)-32 \pi \big(\big(r^2 T^{0\textsf{(Cor)}}_{0}-1\big)
e^{r^2 X}-2 r^2 X+1\big)\big\}+\frac{1}{\pi }\{\alpha \big(-2 L
r^2\\\label{44}&+e^{r^2 X}-1\big)\}\bigg],\\\nonumber
\check{p_r}&=\frac{e^{-r^2 X}}{8 r^2} \bigg[\frac{1}{\psi ^2+12 \pi
\psi +32 \pi ^2}\{-2 \psi \big(e^{r^2 X} \big(r^2
(T^{0\textsf{(Cor)}}_{0}+3 T^{1\textsf{(Cor)}}_{1})+2\big)-2 \big(3
L r^2\\\nonumber&+r^2 X+1\big)\big)-32 \pi \big(-2 L r^2+\big(r^2
T^{1\textsf{(Cor)}}_{1}+1\big) e^{r^2
X}-1\big)\}+\frac{1}{\pi}\{\alpha \big(2 L r^2+1\\\label{46}&-e^{r^2
X}\big)\}\bigg],
\\\nonumber
\check{p_t}&=\frac{1}{16} e^{-r^2 X} \bigg[\{\alpha  \big(2 r
\sqrt{X} \big(2 L^3 r^2+L^2 \big(-4 \alpha +2 r^2 X \big(2 \alpha
+e^{r^2 X}+2\big)+1\big)\\\nonumber&+2 L X \big(\alpha +r^2 X
\big((2 \alpha +1) e^{r^2 X}-1\big)-2 \alpha  e^{r^2 X}+3\big)+X^2
\big(2 \alpha e^{r^2 X}-1\big)\big)\\\nonumber&-\sqrt{\pi } (L+X)
e^{r^2 X} \text{Erf}\big(r \sqrt{X}\big) \big(2 L^2 r^2+L \big(-4
\alpha +2 r^2 (2 \alpha X+X)+1\big)\\\nonumber&+2 \alpha
X+X\big)\big)\}\{\pi \big(\sqrt{\pi } \alpha (L+X) e^{r^2 X}
\text{Erf}\big(r \sqrt{X}\big)-2 r \sqrt{X} \big(\alpha  L+X
\big(\alpha  e^{r^2 X}\\\nonumber&-1\big)\big)\big)\}^{-1}-\{4
\big(\psi \big(e^{r^2 X} \big(r^2 (T^{0\textsf{(Cor)}}_{0}+3
T^{1\textsf{(Cor)}}_{1})+2\big)-2 \big(3 L r^2+r^2
X+1\big)\big)\\\label{46}&+16 \pi  \big(-2 L r^2+\big(r^2
T^{1\textsf{(Cor)}}_{1}+1\big) e^{r^2 X}-1\big)\big)\}\{\big(\psi
^2+12 \pi \psi +32 \pi ^2\big) r^2\}^{-1}\bigg],
\\\nonumber
\check{\Delta}&=\{\alpha  e^{-r^2 X} \big(-\sqrt{\pi } (L+X) e^{r^2
X} \text{Erf}\big(r \sqrt{X}\big) \big(-2 L^2 r^4-L \big(2 r^4 (2
\alpha  X+X)\\\nonumber&+r^2\big)-r^2 (2 \alpha  X+X)+2 \alpha
\big(e^{r^2 X}-1\big)\big)-2 r \sqrt{X} \big(2 L^3 r^4+L^2
\big(r^2+2 r^4 X\\\nonumber&\times \big(2 \alpha +e^{r^2
X}+2\big)\big)+2 L \big(\alpha +(\alpha +1) r^2 X-\alpha  e^{r^2
X}+r^4 X^2 \big((2 \alpha +1) e^{r^2 X}\\\nonumber&-1\big)\big)+X
\big(-2 \alpha  e^{2 r^2 X}+2 e^{r^2 X} \big(\alpha +\alpha  r^2
X+1\big)+r^2 (-X)-2\big)\big)\big)\}\\\label{56}&\times\{16 \pi  r^2
\big(\sqrt{\pi } \alpha (L+X) e^{r^2 X} \text{Erf}\big(r
\sqrt{X}\big)-2 r \sqrt{X} \big(\alpha L+X \big(\alpha  e^{r^2
X}-1\big)\big)\big)\}^{-1}.
\end{align}

\subsection{The Second Solution}

The second anisotropic solution is computed by utilizing a
density-like constraint, i.e.,
\begin{equation}\label{48}
\widetilde{\rho}+T^{0\textsf{(Cor)}}_{0}=\delta^{0}_{0}.
\end{equation}
Making use of Eq.\eqref{18} along with \eqref{21} and \eqref{22}, we
have
\begin{align}\label{49}
h^\ast&=\frac{(L+\alpha  X) \ln \big(\alpha  \big(e^{r^2
X}-1\big)+1\big)-\alpha  r^2 X (L+X)}{(\alpha-1) \alpha
X},\\\label{50} k^\ast&=1-e^{-r^2 X}.
\end{align}
In this solution, the matching conditions are
\begin{align}\label{51}
\ln\bigg(1-\frac{ \mathbb{M}_o}{\textsf{R}}\bigg)&=\frac{(L+\alpha
X) \ln \big(\alpha  \big(e^{\textsf{R}^2 X}-1\big)+1\big)-\alpha
\textsf{R}^2 X (L+X)}{(\alpha -1) X}+P+L \textsf{R}^2,\\\label{52}
1-\frac{ \mathbb{M}_o}{\textsf{R}}&=\alpha -(\alpha -1)
e^{-\textsf{R}^2 X}.
\end{align}
Finally, the matter variables for solution II are as follows
\begin{align}\nonumber 
\check{\rho}&=\frac{e^{-r^2 X}}{8 r^2} \bigg[\frac{1}{\psi ^2+12 \pi
\psi +32 \pi ^2}\{-2 \psi \big(-2 L r^2+e^{r^2 X} \big(r^2 (3
T^{0\textsf{(Cor)}}_{0}+T^{1\textsf{(Cor)}}_{1})\\\nonumber&-2\big)-6
r^2 X+2\big)-32 \pi \big(\big(r^2 T^{0\textsf{(Cor)}}_{0}-1\big)
e^{r^2 X}-2 r^2 X+1\big)\}-\frac{1}{\pi }\{\alpha \big(2 r^2
X\\\label{53}&+e^{r^2 X}-1\big)\}\bigg],\\\nonumber \check{p_r}&=
\frac{e^{-r^2 X}}{8 r^2} \bigg[\frac{1}{\psi ^2+12 \pi \psi +32 \pi
^2}\{-2 \psi \big(e^{r^2 X} \big(r^2 (T^{0\textsf{(Cor)}}_{0}+3
T^{1\textsf{(Cor)}}_{1})+2\big)-2 \big(3 L\\\nonumber&\times r^2+r^2
X+1\big)\big)-32 \pi \big(-2 L r^2+\big(r^2
T^{1\textsf{(Cor)}}_{1}+1\big) e^{r^2
X}-1\big)\}+\frac{1}{\pi}\{\alpha \big(2 r^2 X\\\label{54}&+e^{r^2
X}-1\big)\}\bigg],
\\\nonumber
\check{p_t}&=\frac{1}{8} e^{-r^2 X} \bigg[\{-2 \psi \big(e^{r^2 X}
\big(r^2 (T^{0\textsf{(Cor)}}_{0}+3
T^{1\textsf{(Cor)}}_{1})+2\big)-2 \big(3 L r^2+r^2
X+1\big)\big)\\\nonumber&-32 \pi \big(-2 L r^2+\big(r^2
T^{1\textsf{(Cor)}}_{1}+1\big) e^{r^2 X}-1\big)\}\{\big(\psi ^2+12
\pi \psi +32 \pi ^2\big) r^2\}^{-1}\\\nonumber&-\{\alpha \big(L^2
r^2 \big(e^{r^2 X}-1\big)+L r^2 X \big(e^{r^2 X}-2\big)+X \big(3
\alpha +r^2 X \big(2 \alpha \big(e^{r^2
X}-1\big)\\\label{55}&+1\big)-3 \alpha e^{r^2 X}-3\big)\big)\}\{\pi
\big(\alpha  \big(e^{r^2 X}-1\big)+1\big)\}^{-1}\bigg].
\end{align}
The anisotropic expression for this solution is
\begin{align}\nonumber
\check{\Delta}&=\{\alpha  e^{-r^2 X} \big(\alpha -L^2 r^4+e^{r^2 X}
\big(L^2 r^4+L r^4 X+\alpha  \big(2 r^4 X^2-r^2
X-2\big)+1\big)\\\nonumber&-2 L r^4 X-2 \alpha  r^4 X^2+r^4
X^2+\alpha r^2 X+\alpha  e^{2 r^2 X}-r^2 X-1\big)\}\{8 \pi  r^2
\big(\alpha \big(e^{r^2 X}\\\label{57}&-1\big)+1\big)\}^{-1}.
\end{align}

\section{Essential Characteristics}

This section investigates some feasible and stable features of the
acquired anisotropic solutions. To do so, we consider the model
\eqref{60} and choose the parameters $\psi$ and $\chi$ as -13 and 1,
respectively. For solution I, the positive values of $\psi$
correspond to the acceptable behavior of matter variables and energy
constraints. However, all the physical tests fail to check the
stability of the first solution (not plotted here). This leads us to
select the negative value of $\psi$ as -13 for both solutions. This
value of coupling parameter shows acceptable behavior of state
variables, energy conditions and stability criteria corresponding to
both solutions. We can thus deduce that the negative value of $\psi$
provides compatible solutions, while positive values produce the
encounter behavior. Moreover, this leads to the fact that the
positive $\psi$ does not yield consistent results and hence cannot
discuss the self-gravitating bodies.

The constant terms (L and P) are interpreted from \eqref{36} and
\eqref{36a} whereas X is selected from \eqref{41}. The feasibility
of the compact structure is checked through the behavior of
effective matter distributions. The effective energy density and
pressure ingredients should be maximum, finite as well as positive
near the center and must decrease with an increment in $r$. For
solution I, the plots of density and anisotropic pressures in Figure
\textbf{1} illustrate the maximal trend near the center and then
show monotonic decreasing behavior towards the boundary with $r$. It
is also seen that the tangential/radial pressures at the surface of
star are zero. The last plot of Figure \textbf{1} shows that
anisotropy disappears at the center while it becomes paramount on
reaching the boundary. One can also examine that anisotropy is zero
at the center for all values of the decoupling parameter, whereas at
the stellar surface, the anisotropy increases by increasing $\alpha$
which assures that the extra source generates anisotropy in the
system. To examine the internal realistic fluid of the compact
object, some bounds are imposed on the EMT, known as energy
conditions. These constraints guarantee the existence of normal
matter and viability of the developed solutions. Four energy
conditions, i.e., weak (WEC), null (NEC), dominant (DEC) and strong
(SEC) for anisotropic configuration are specified as
\begin{align}\nonumber
\text{NEC:}\quad&\check{\rho}+\check{p_r}\geq0,\quad\check{\rho}+\check{p_t}\geq0,
\\\nonumber \text{WEC:}\quad &\check{\rho}+\check{p_t}\geq0,\quad\check{\rho}\geq0,
\quad \check{\rho}+\check{p_r}\geq0,
\\\nonumber\text{DEC:}\quad &\check{\rho}-\check{p_t}\geq0,\quad
\check{\rho}-\check{p_r}\geq0,
\\\label{59}\text{SEC:}\quad
&\check{\rho}+\check{p_r}+2\check{p_t}\geq0.
\end{align}
\begin{figure}\center
\epsfig{file=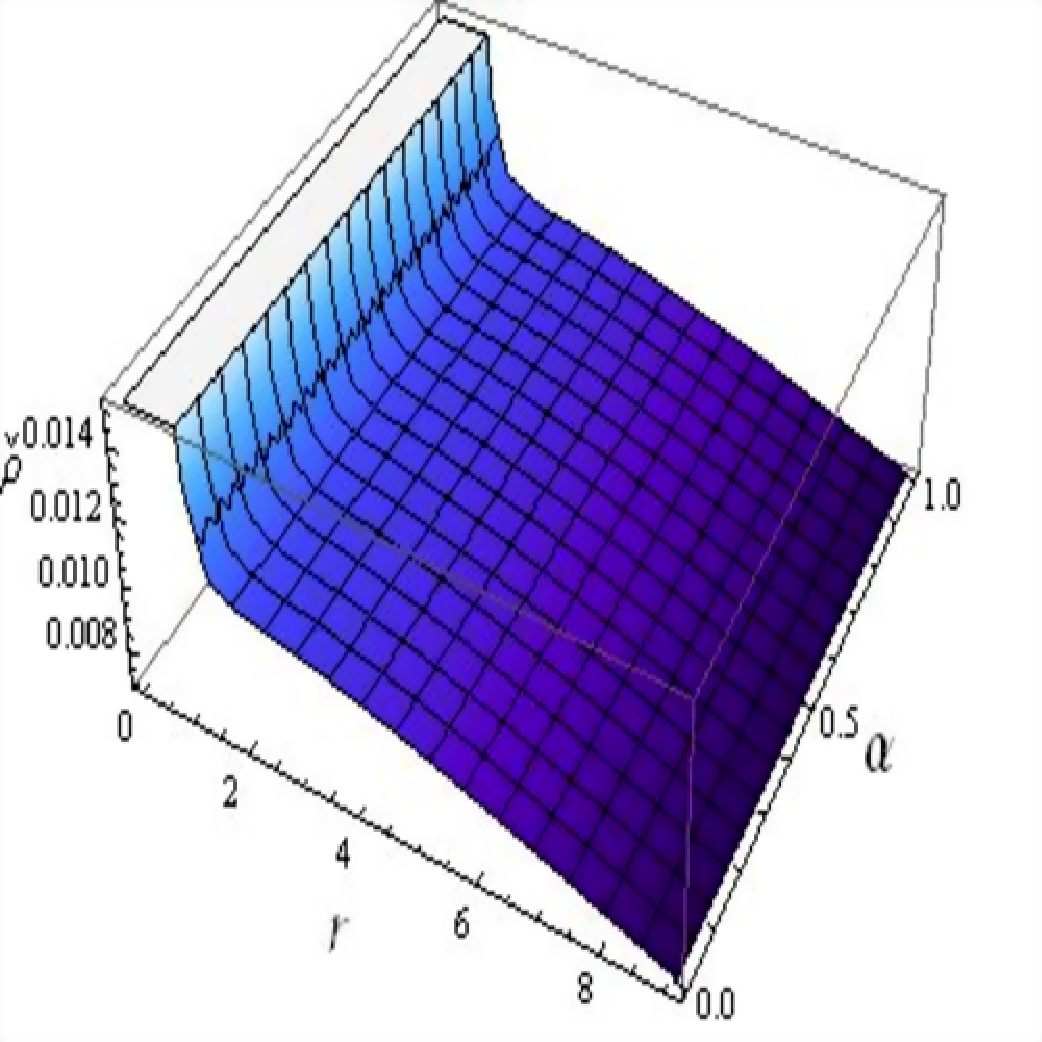,width=0.4\linewidth}\epsfig{file=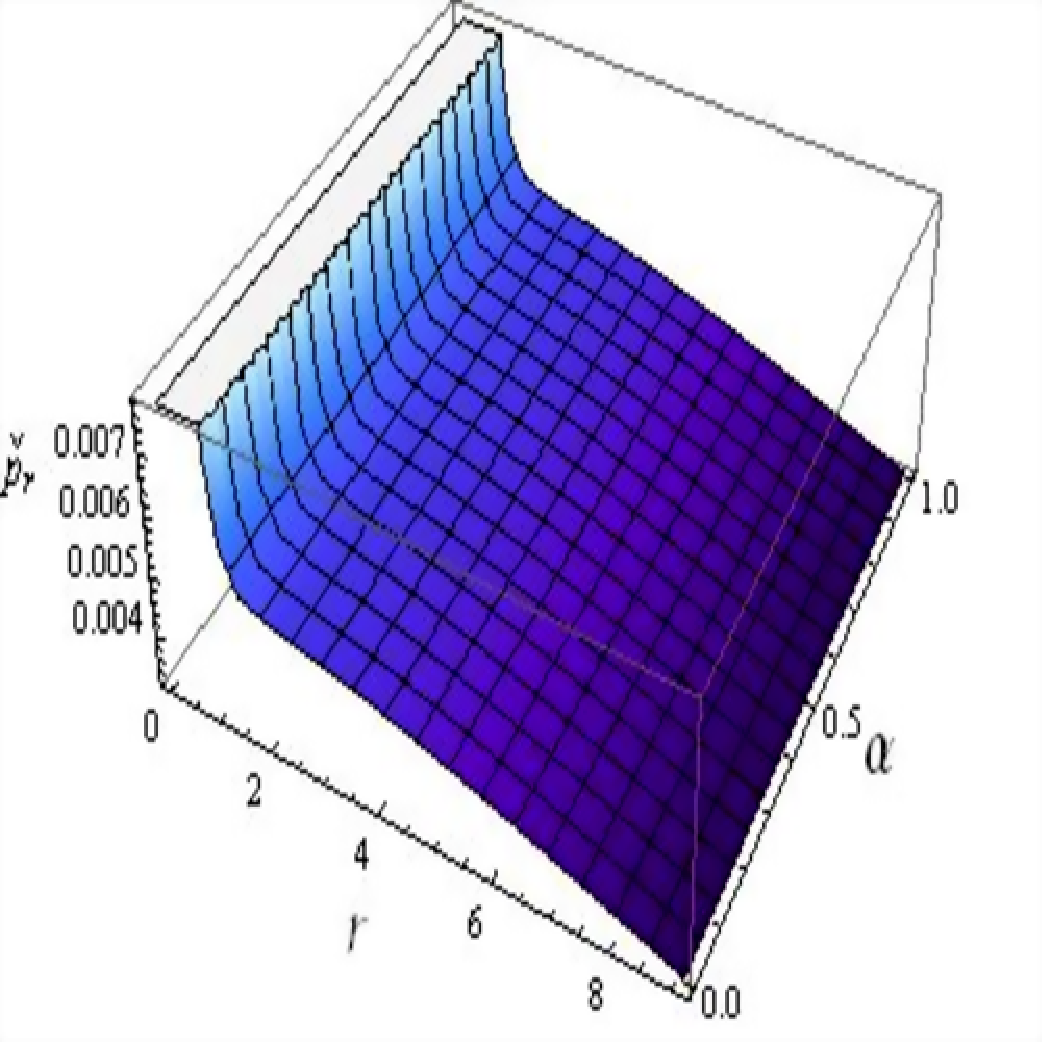,width=0.4\linewidth}
\epsfig{file=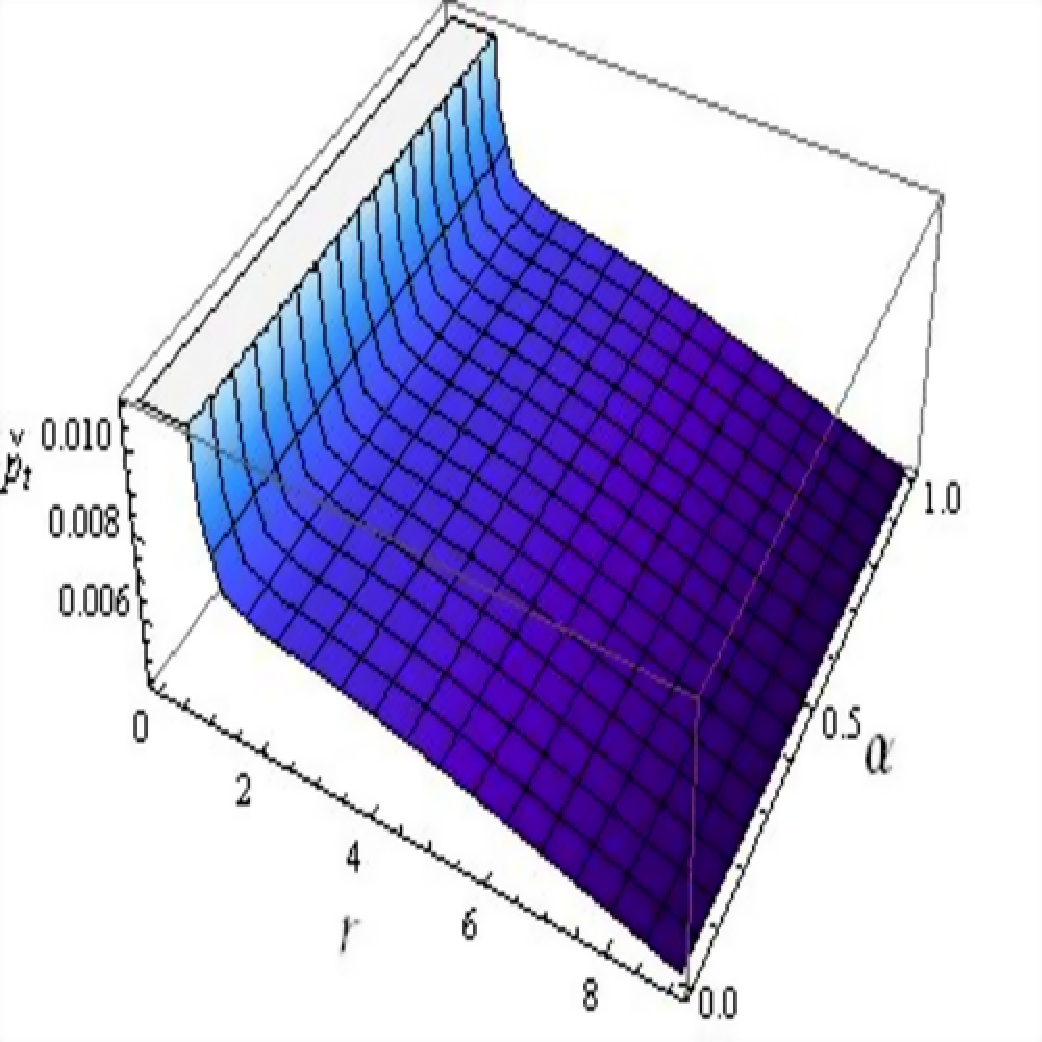,width=0.4\linewidth}\epsfig{file=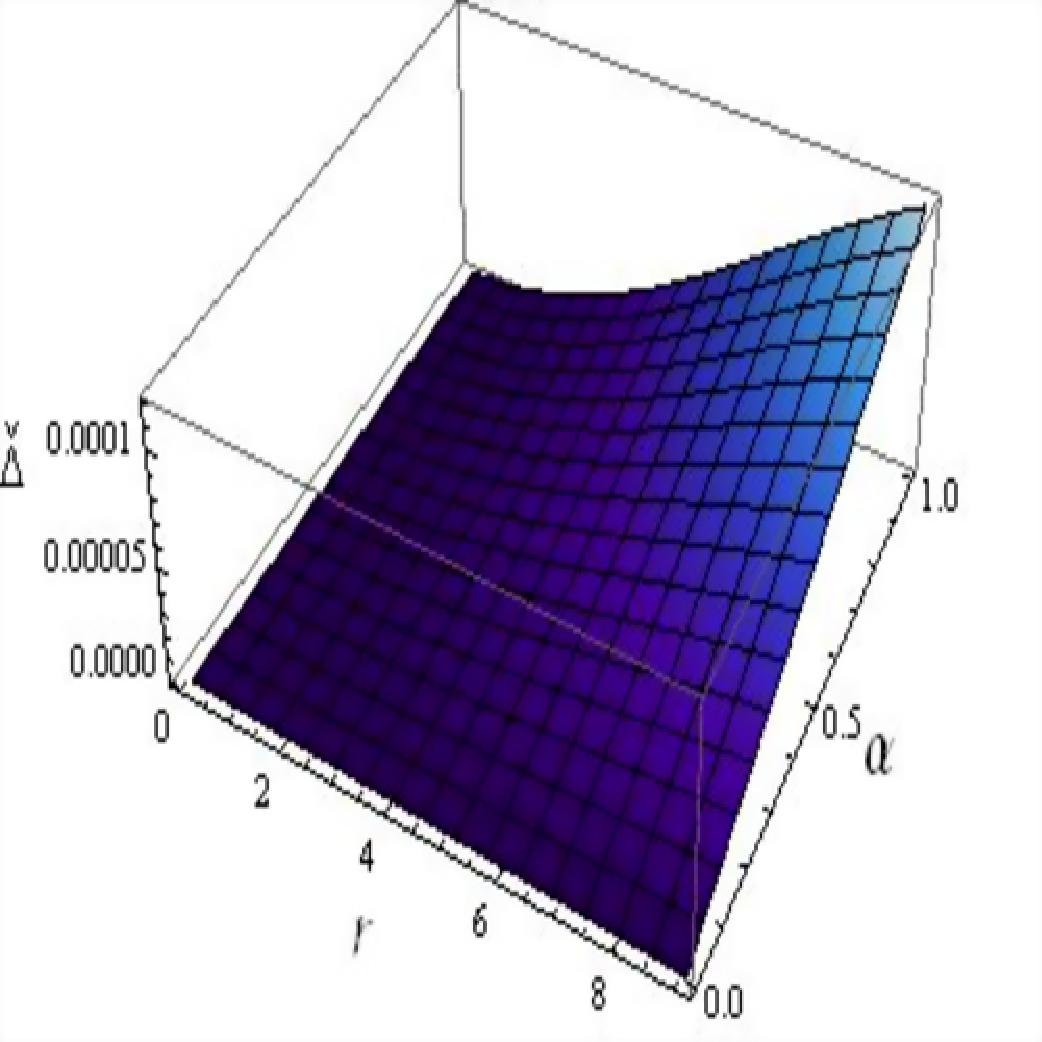,width=0.4\linewidth}
\caption{Analysis of $\check{\rho},\check{p_r},\check{p_t}$ (density
and pressure components) and $\check{\Delta}$ (anisotropy) versus
$r$ and $\alpha$ for the solution I.}
\end{figure}
\begin{figure}\center
\epsfig{file=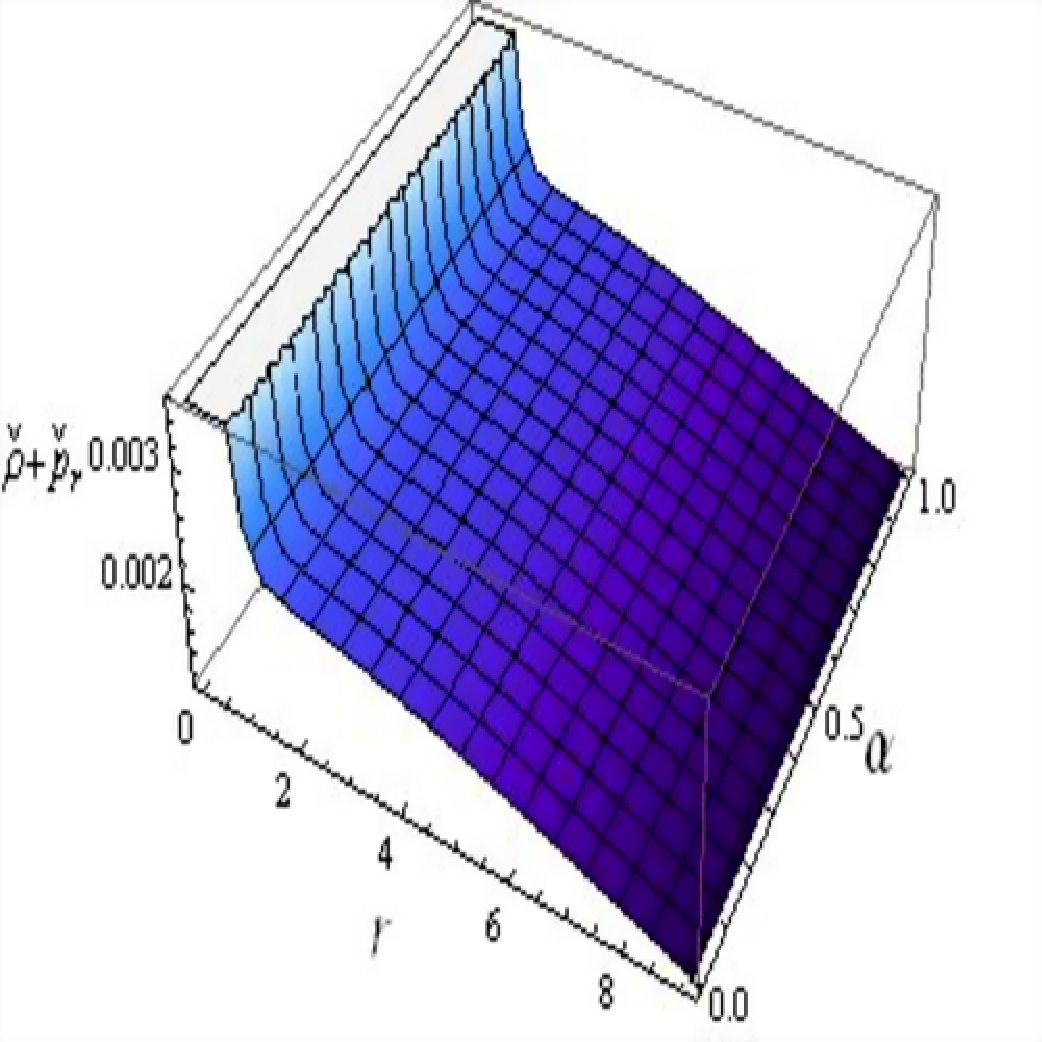,width=0.4\linewidth}\epsfig{file=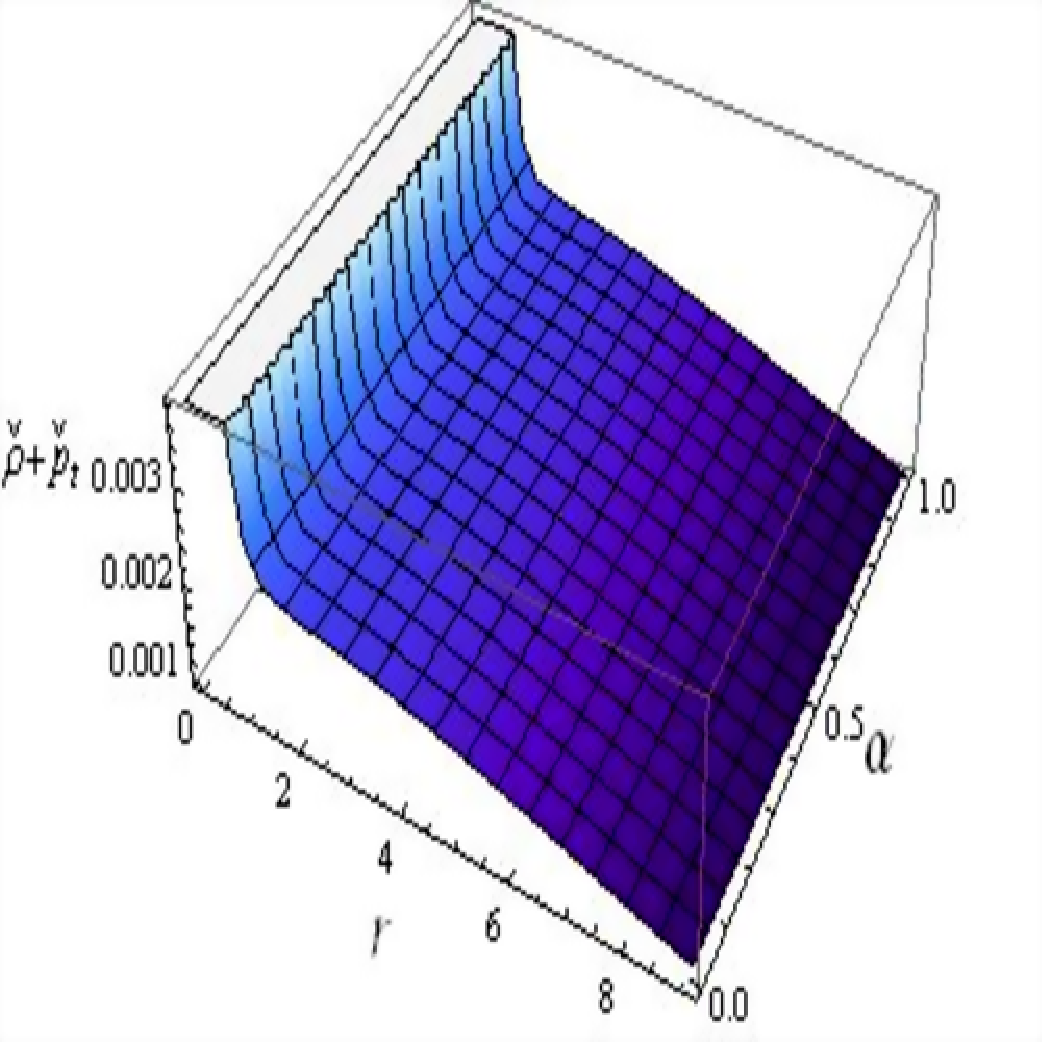,width=0.4\linewidth}
\epsfig{file=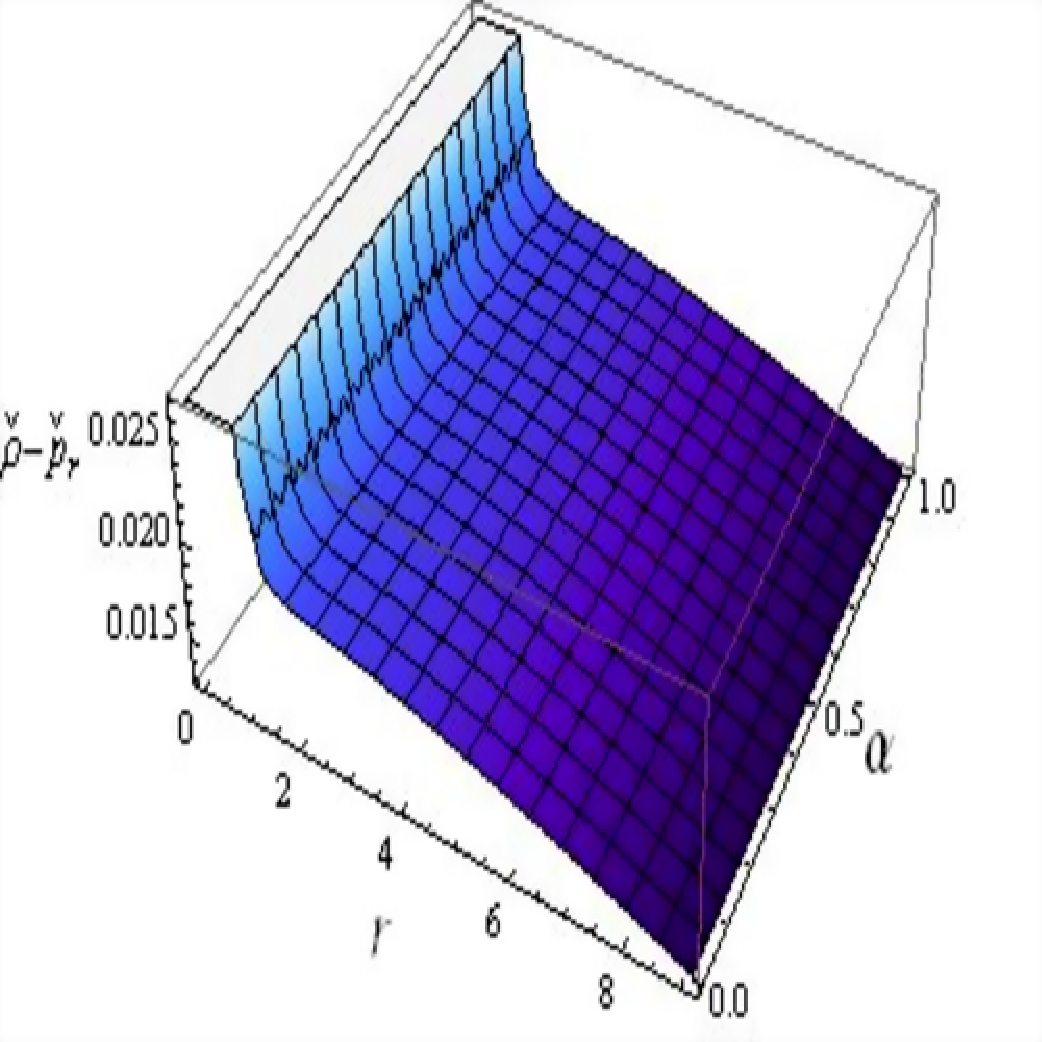,width=0.4\linewidth}\epsfig{file=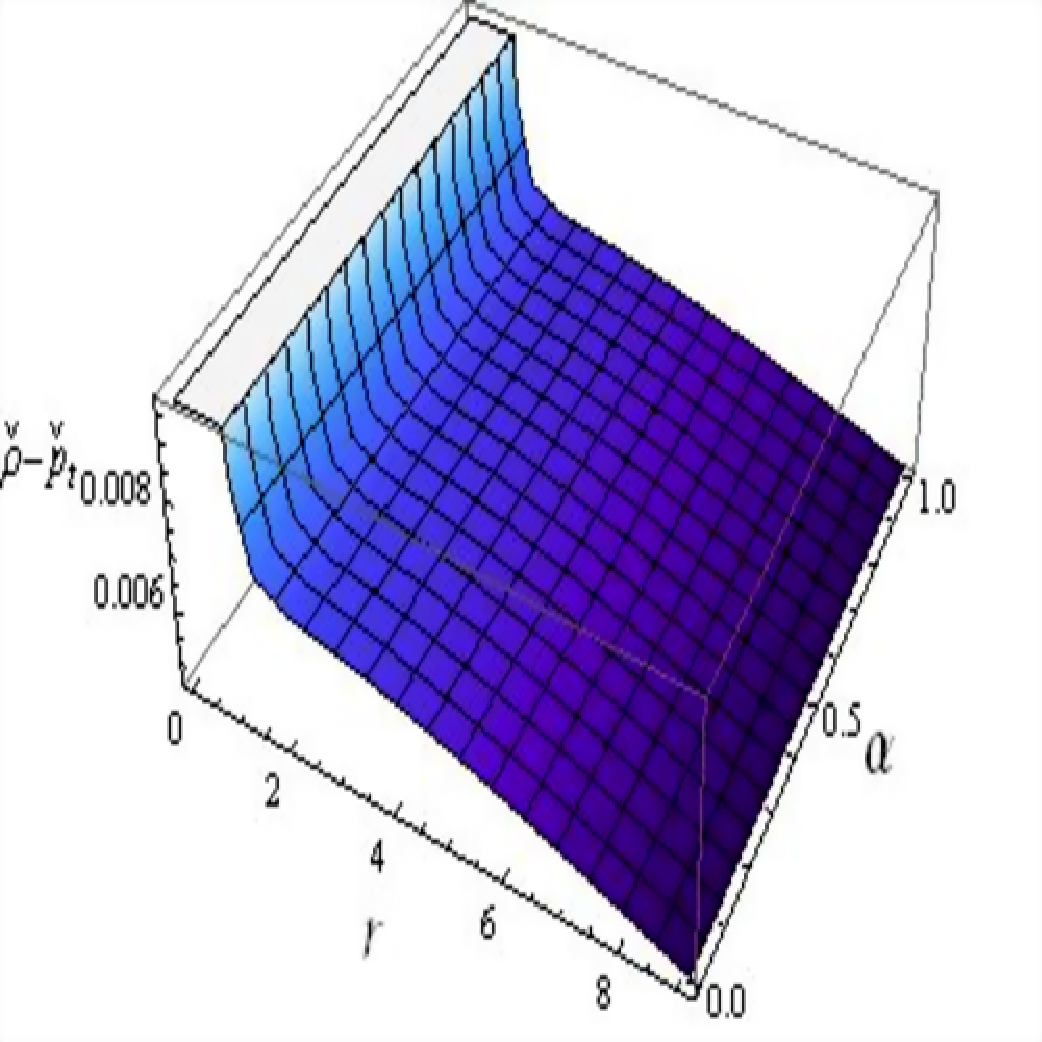,width=0.4\linewidth}
\epsfig{file=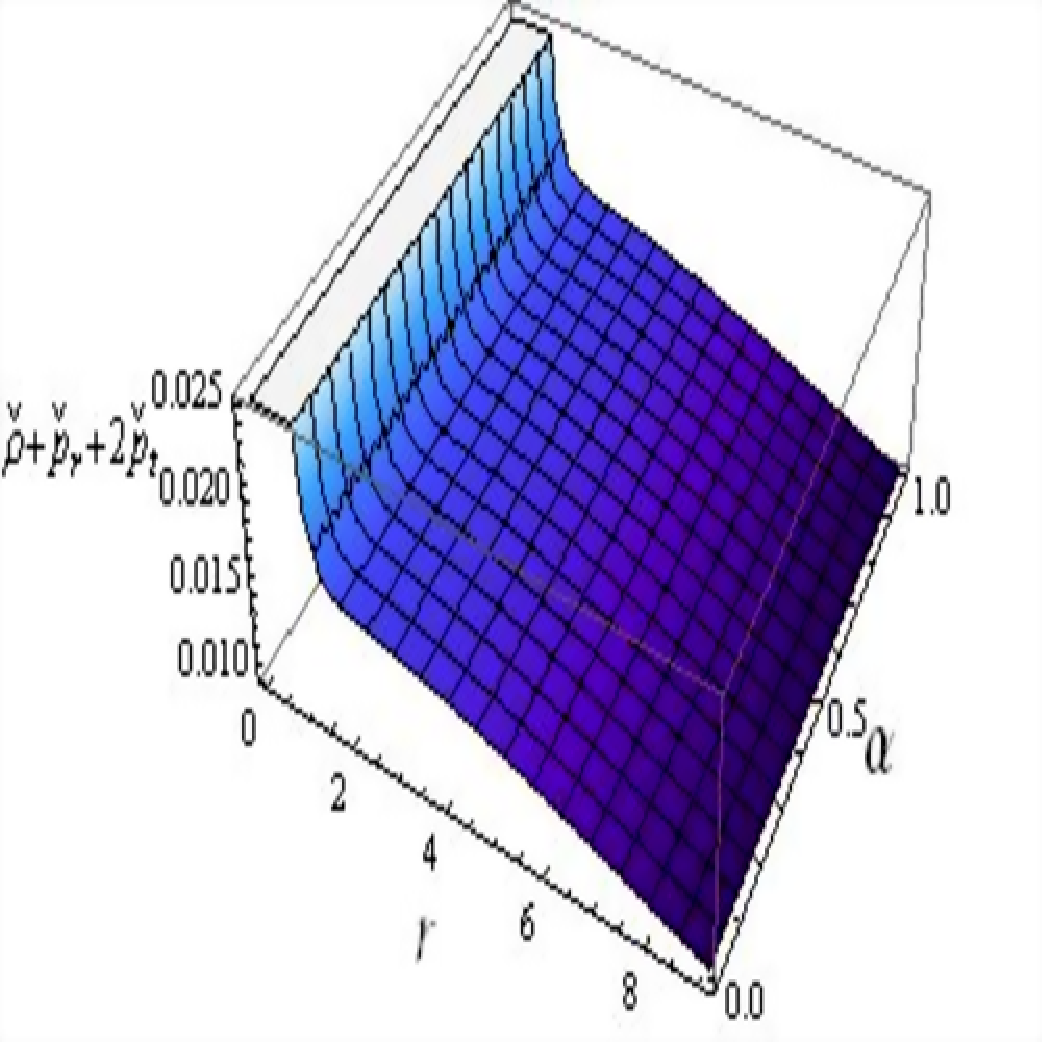,width=0.4\linewidth} \caption{Analysis of energy
constraints for the solution I.}
\end{figure}

Stability of a compact structure is a crucial factor to examine the
physical acceptability of the derived models. There are different
methods to gauge the stable structure of celestial object. One of
the techniques is causality condition, which states that the speed
of light must always be faster than the speed of sound. The
components of the speed sound are represented as
\begin{equation}\label{59a}
\nu^{2}_{t}=\frac{d\check{p_{t}}}{d\check{\rho}},\quad\quad\nu^{2}_{r}=\frac{d\check{p_{r}}}{d\check{\rho}},
\end{equation}
where $\nu^2_{t}$ and $\nu^2_{r}$ are the tangential and radial
square speed components, respectively, with $0<\nu^2_{t}<1$ as well
as $0<\nu^2_{r}< 1$ \cite{24ab}. Another way to determine the
stability is proposed by Herrera \cite{24}, i.e., cracking approach
according to which the constituents of sound speed associated with
the stellar system should lie in $|\nu^{2}_{t}-\nu^{2}_r|<1$. The
adiabatic index is an alternative technique which supports the
stable behavior of astronomical objects defined as
\begin{equation}\label{59b}
\check{\Gamma_{r}}=\frac{\check{\rho}+\check{p_{r}}}{\check{p_{r}}}\bigg(\frac{d\check{p_{r}}}{d\check{\rho}}\bigg),\quad
\check{\Gamma_{t}}=\frac{\check{\rho}+\check{p_{t}}}{\check{p_{t}}}\bigg(\frac{d\check{p_{t}}}{d\check{\rho}}\bigg).
\end{equation}
The astrophysical structure shows the stable region if radial as
well as tangential part of adiabatic index is greater than
$\frac{4}{3}$ \cite{25}. Figure \textbf{2} indicates that all the
energy conditions for the first solution comply with the required
limits, so the solution I is viable. It is also clear from Figure
\textbf{3} that the first solution  meets the requirements of all
three stability criteria, which indicates the stability of solution
I.
\begin{figure}\center
\epsfig{file=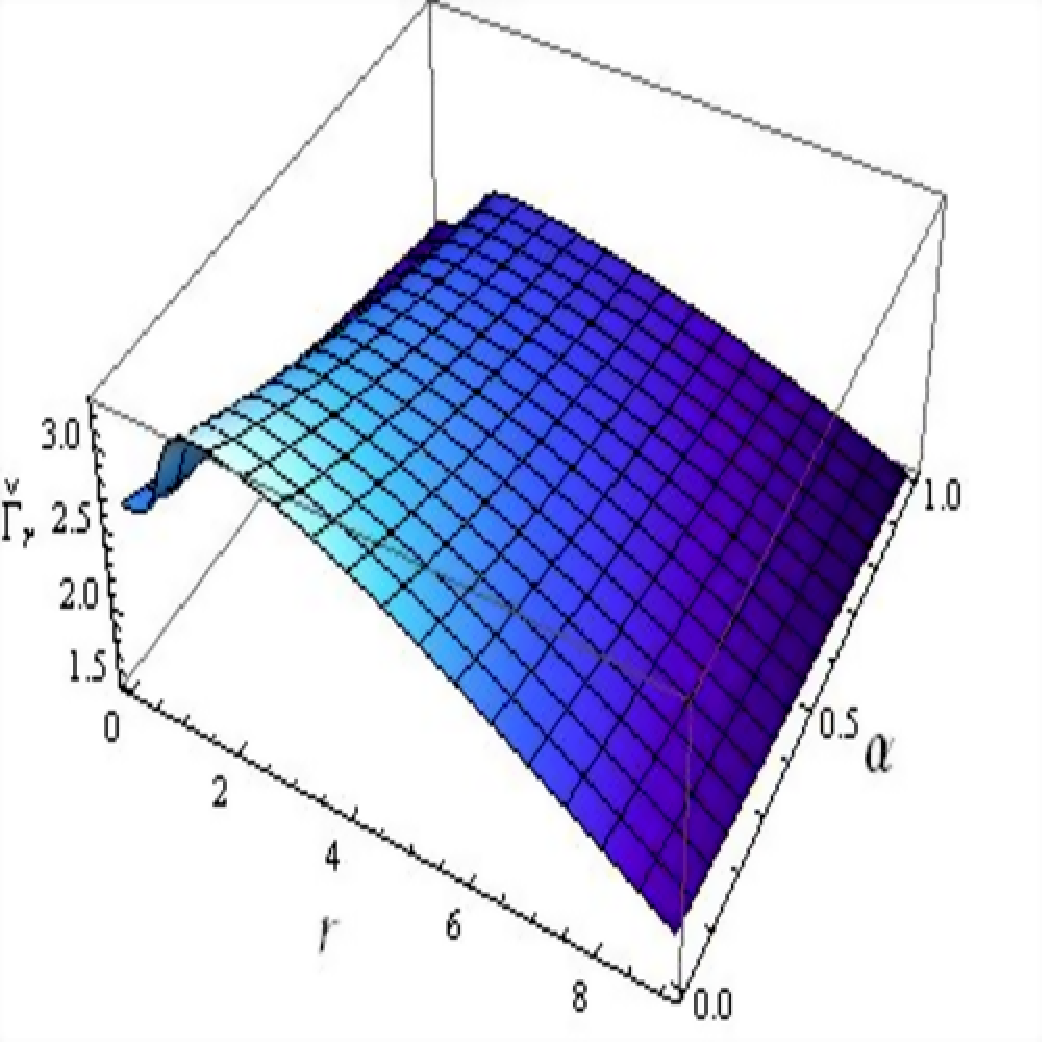,width=0.4\linewidth}\epsfig{file=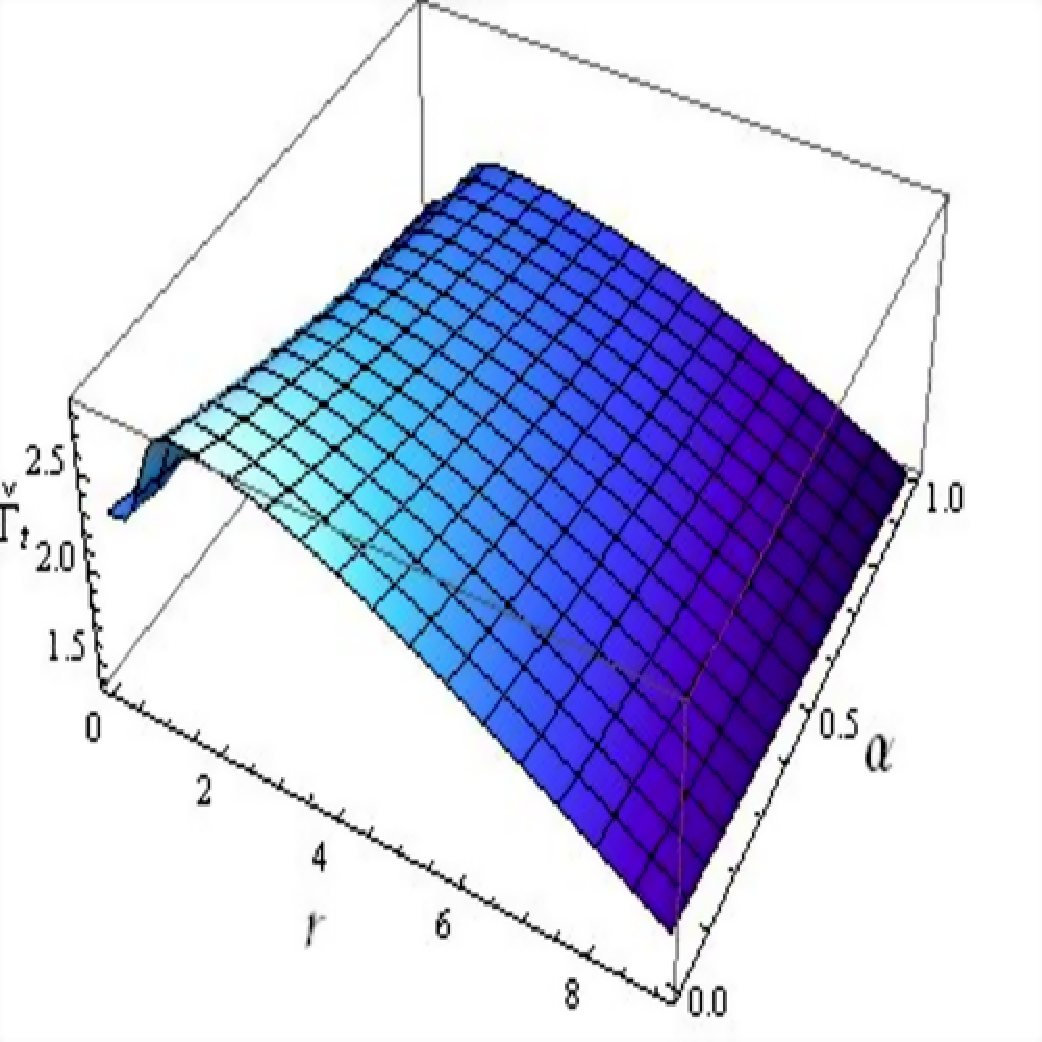,width=0.4\linewidth}
\epsfig{file=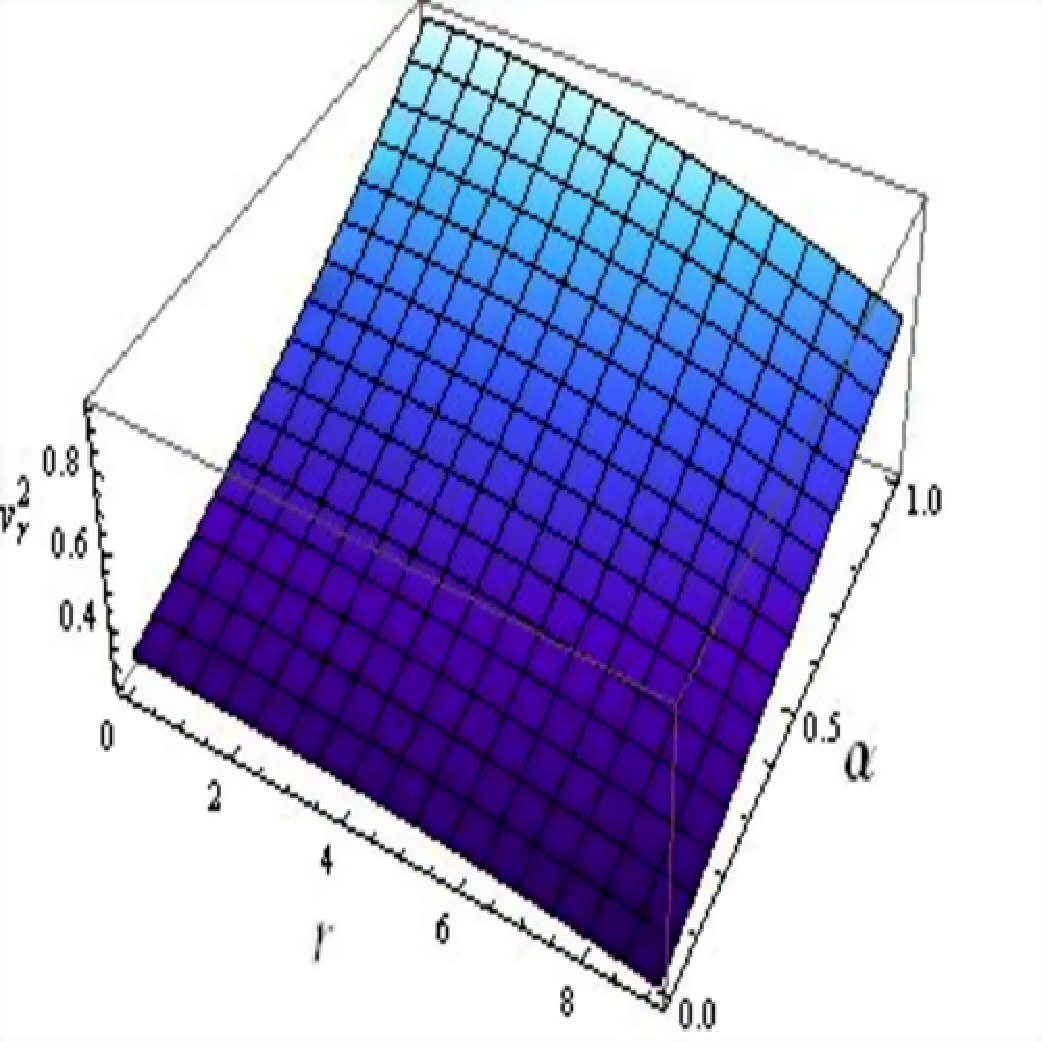,width=0.4\linewidth}\epsfig{file=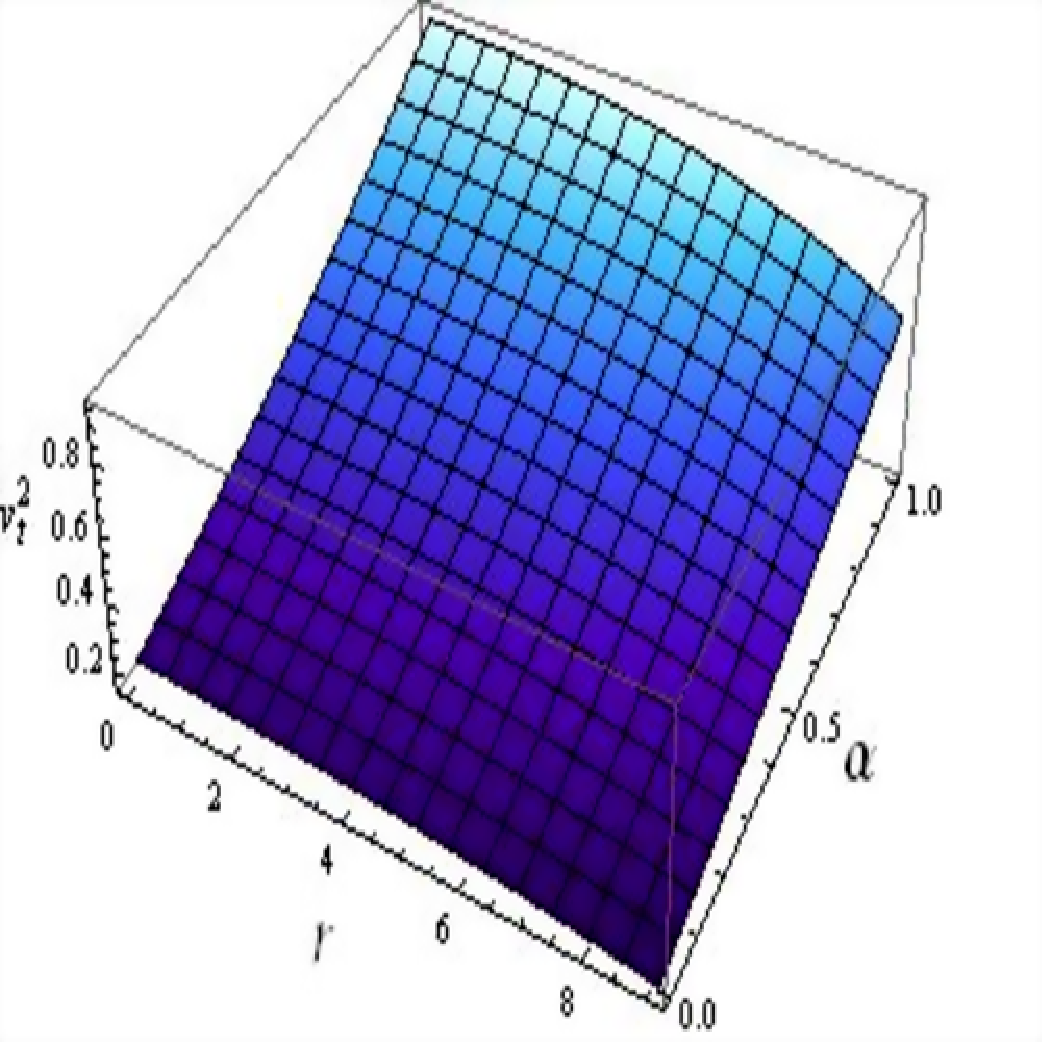,width=0.4\linewidth}
\epsfig{file=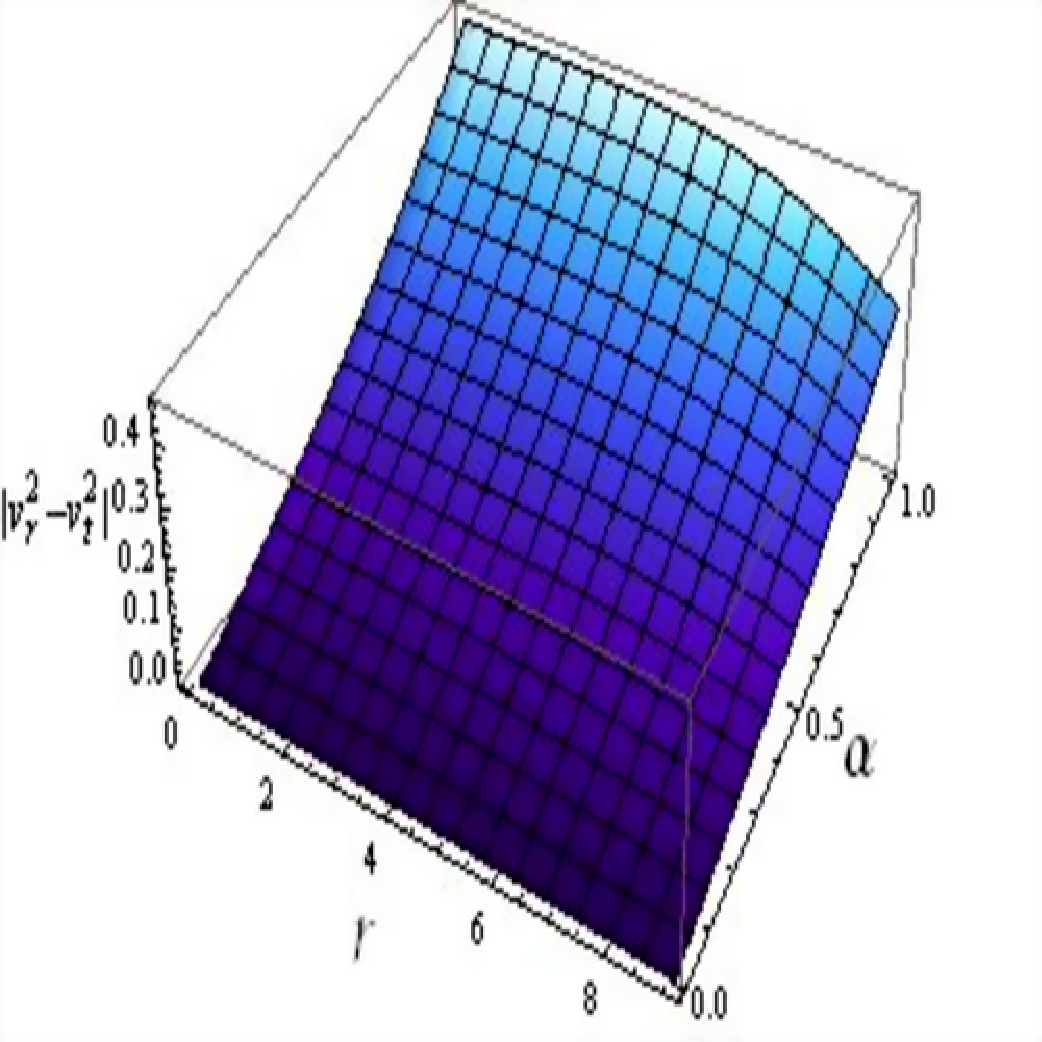,width=0.4\linewidth} \caption{Analysis of
adiabatic index, causality condition and Herrera cracking approach
versus $r$ and $\alpha$ for the solution I.}
\end{figure}

The physical analysis of the second solution is accomplished by
considering the same values as chosen in solution I. The matter
determinants, likewise in solution I, must be finite, maximum and
decreasing with $r$. The energy density, radial/tangential pressures
in Figure \textbf{4} are maximum near the center and decline towards
the boundary as $r$ increases. The graphical representation of
anisotropy in Figure \textbf{4} exhibits zero behavior at the center
and persists this behavior throughout $r$. It is also observed that
for all values of $\alpha$, anisotropy becomes zero at the core
while it displays an increment at the star surface with larger
values of the decoupling parameter. The viability of solution II is
shown in Figure \textbf{5} as energy constraints are satisfied.
Figure \textbf{6} shows the stable behavior of solution II through
both causality condition as well as Herrera cracking approach.
Moreover, the adiabatic index is also validated in the whole domain
since its radial and tangential components lie within the stable
range (Figure \textbf{6}). Hence, solution II demonstrates the
stable behavior according to all the criteria.
\begin{figure}\center
\epsfig{file=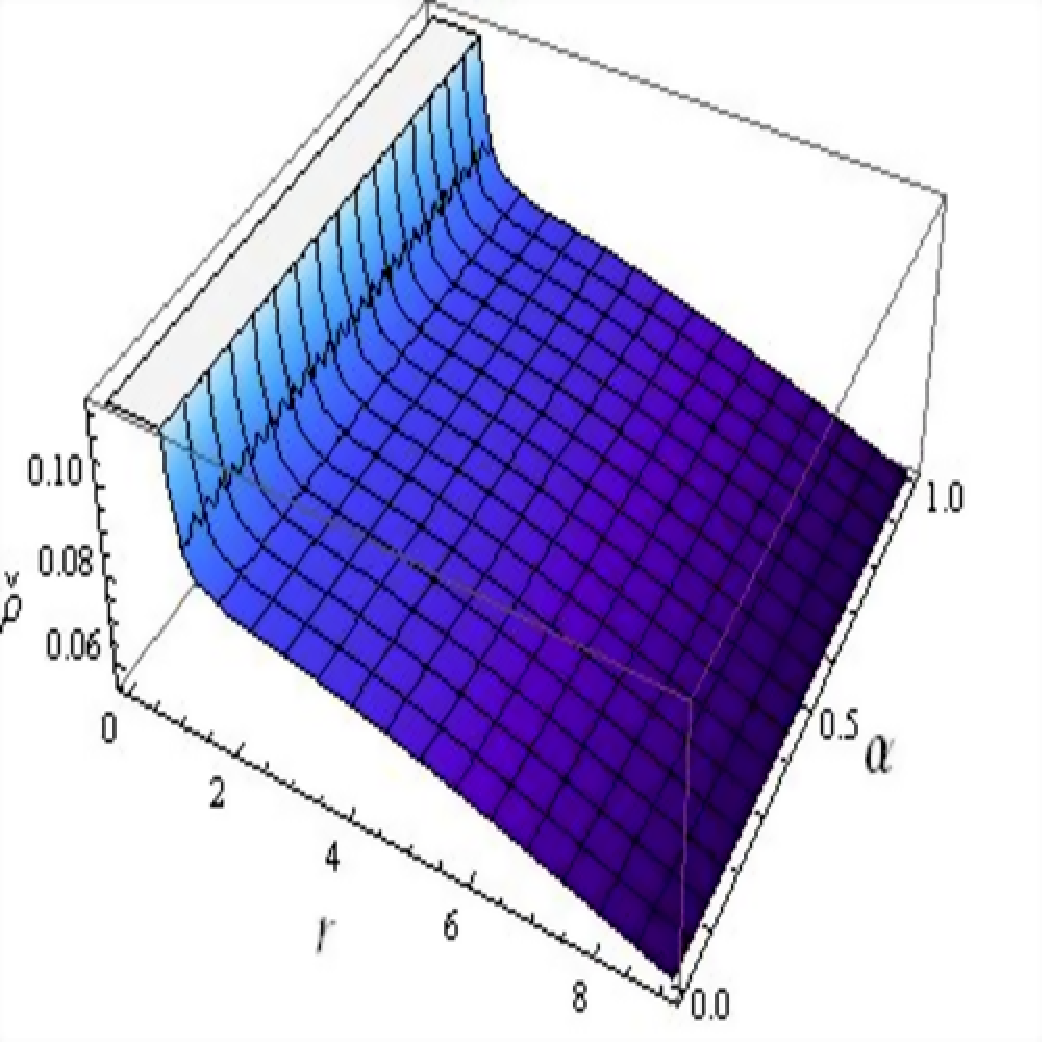,width=0.4\linewidth}\epsfig{file=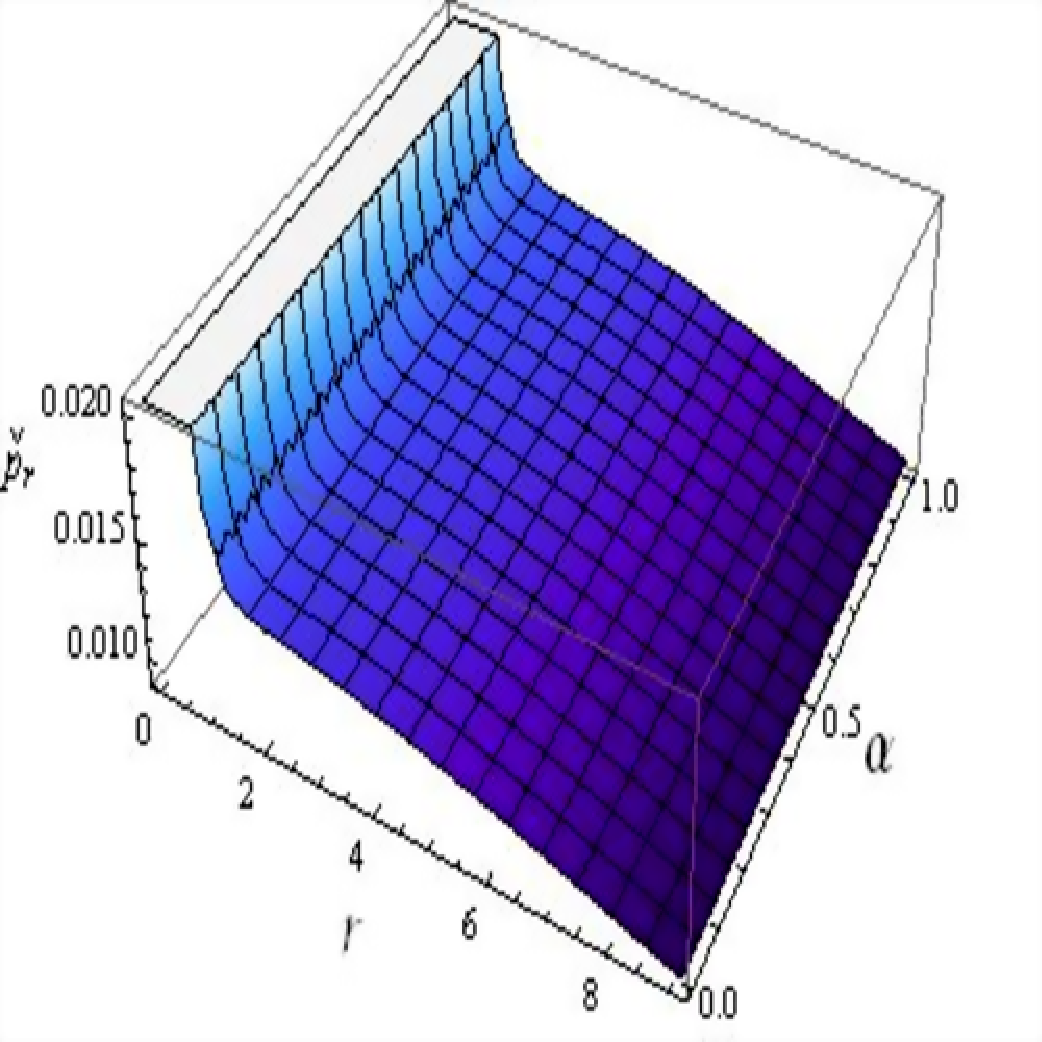,width=0.4\linewidth}
\epsfig{file=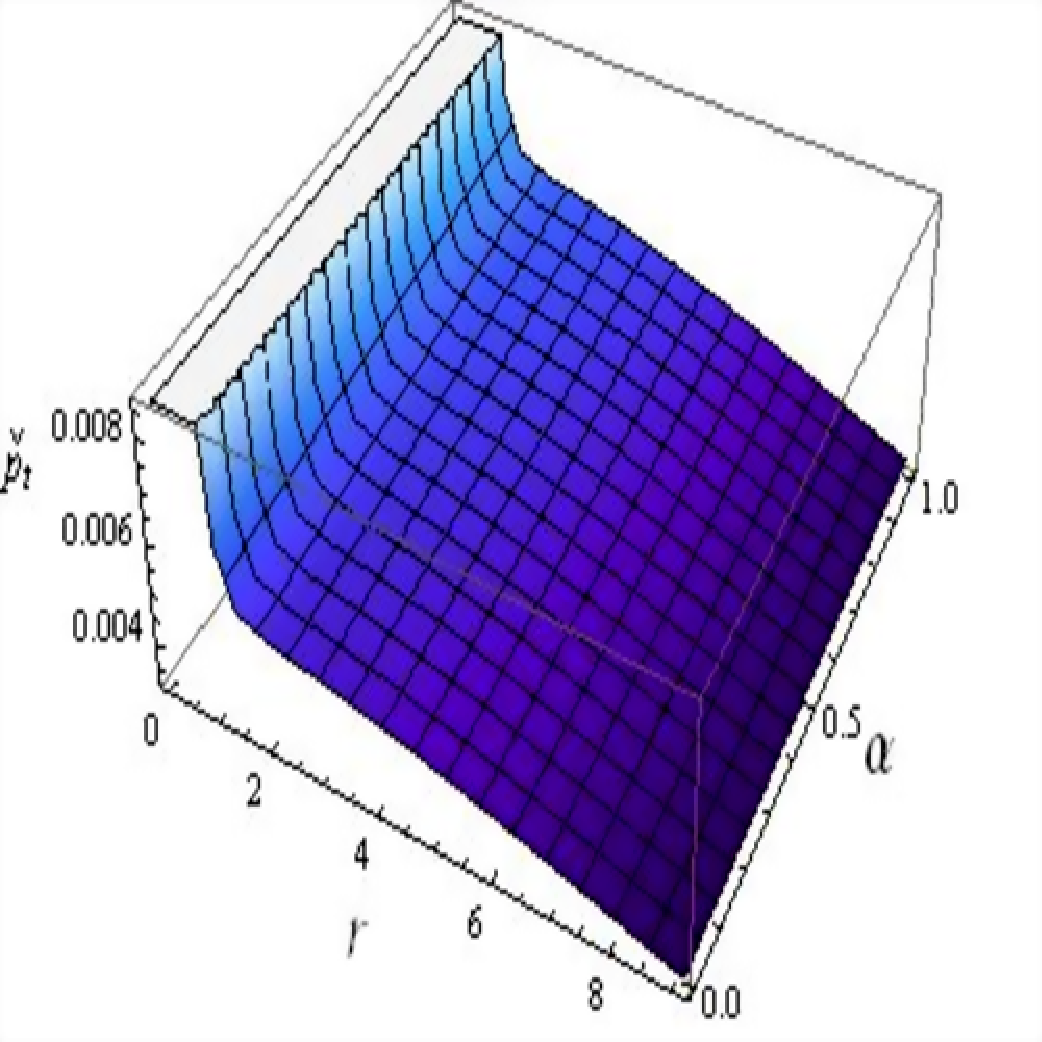,width=0.4\linewidth}\epsfig{file=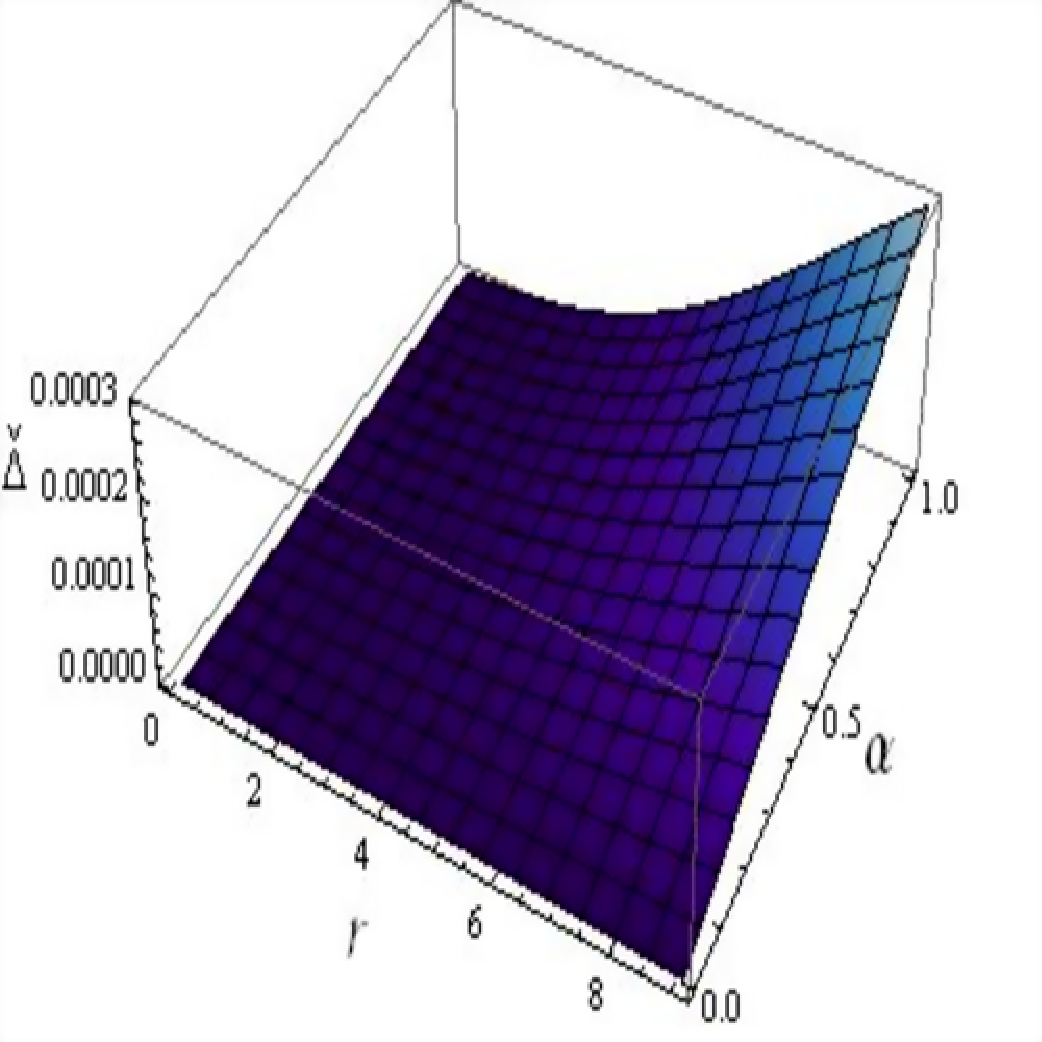,width=0.4\linewidth}
\caption{Analysis of $\check{\rho},\check{p_r},\check{p_t}$ (density
and pressure components) and $\check{\Delta}$ (anisotropy) versus
$r$ and $\alpha$ for the solution II.}
\end{figure}
\begin{figure}\center
\epsfig{file=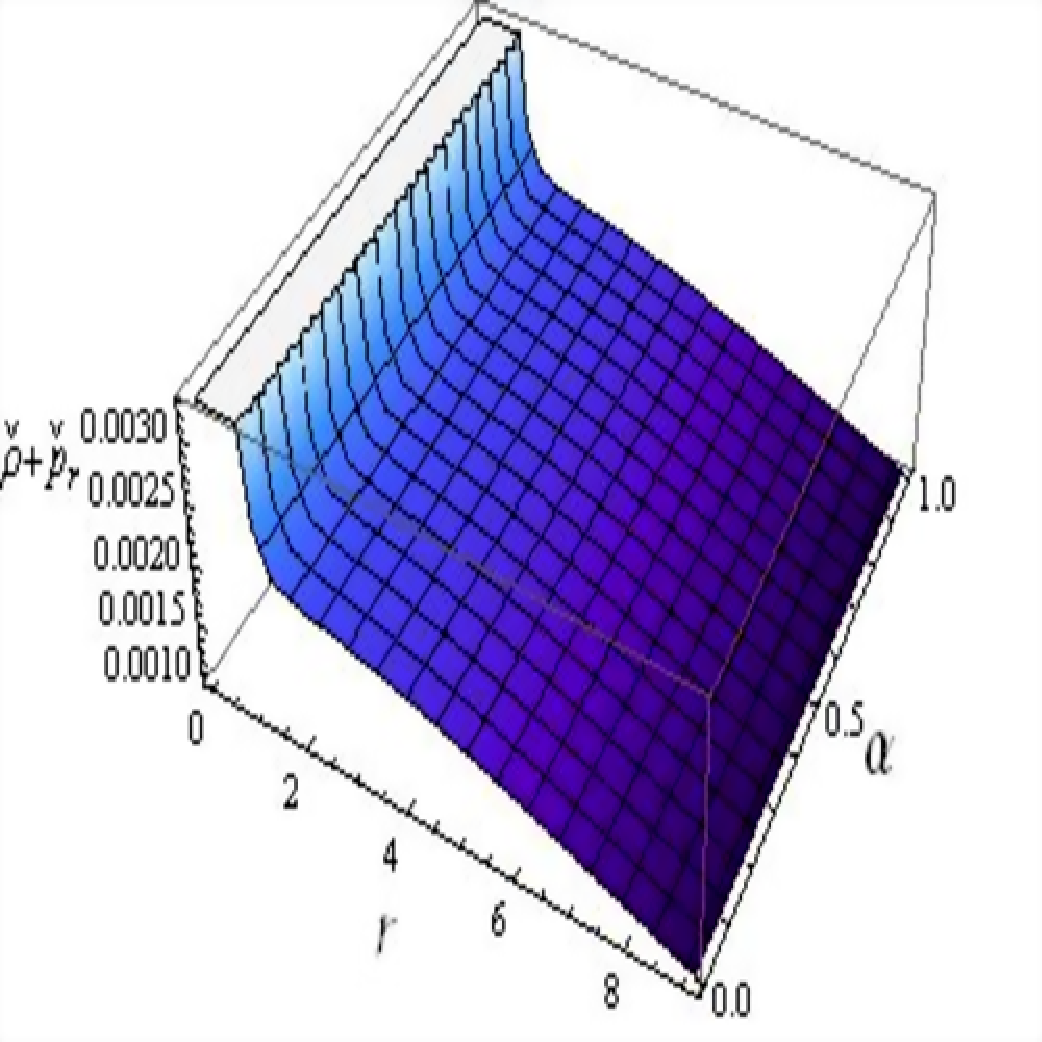,width=0.4\linewidth}\epsfig{file=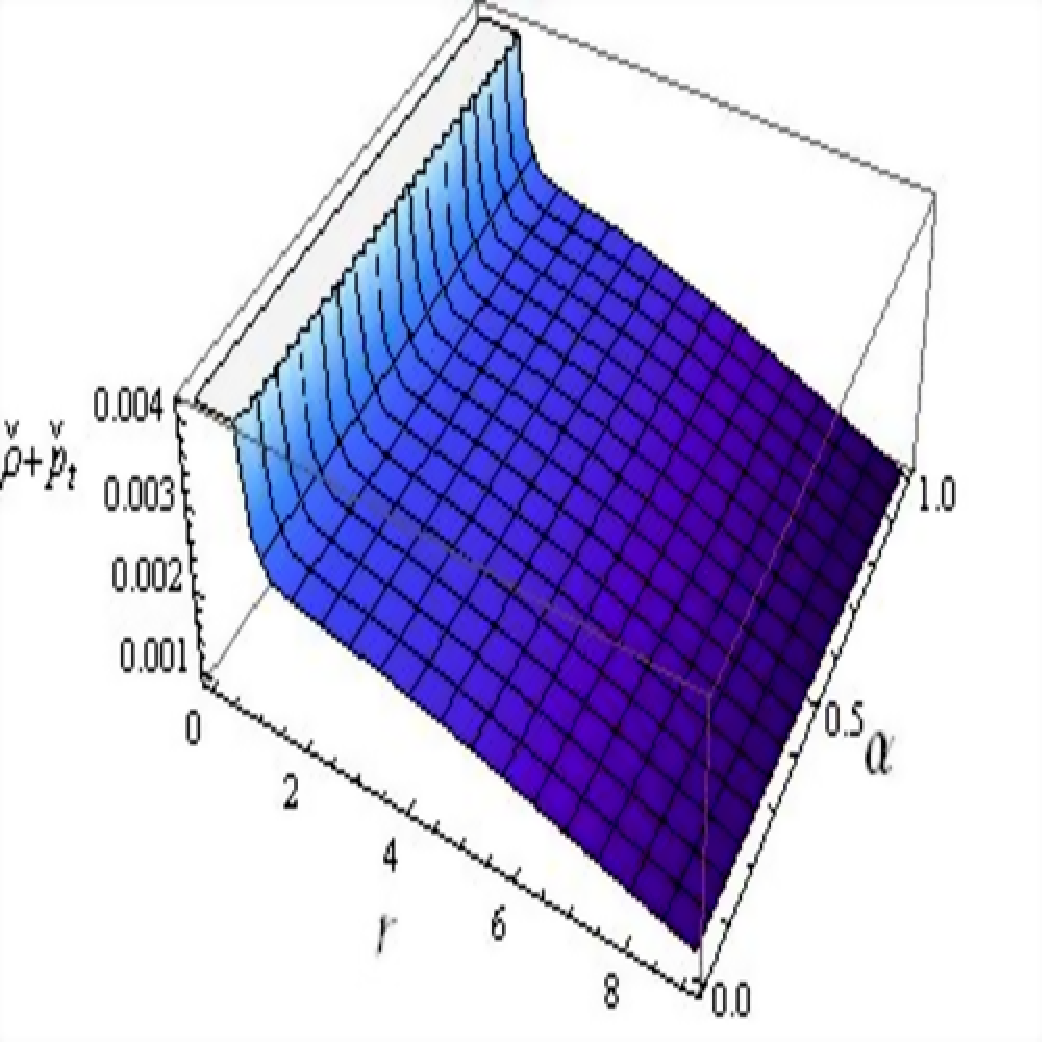,width=0.4\linewidth}
\epsfig{file=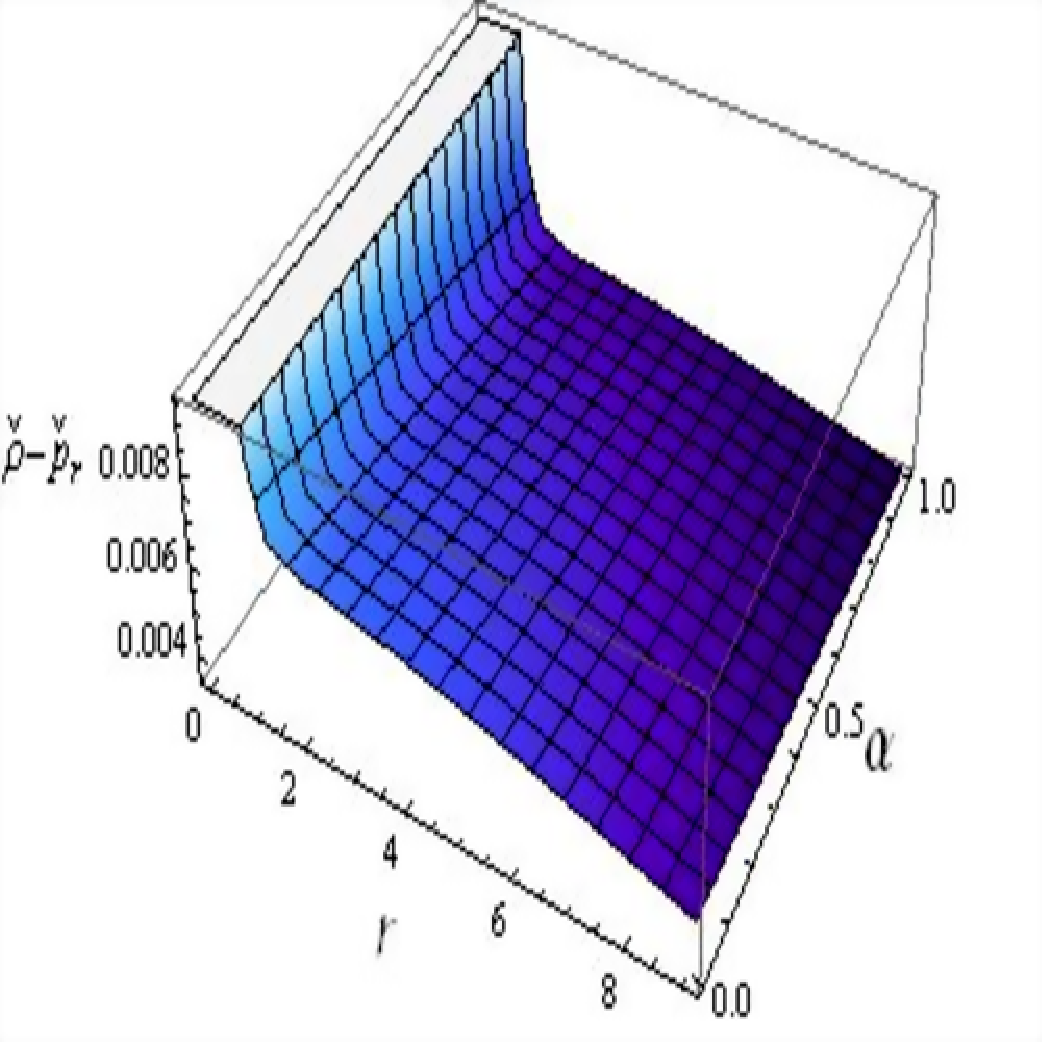,width=0.4\linewidth}\epsfig{file=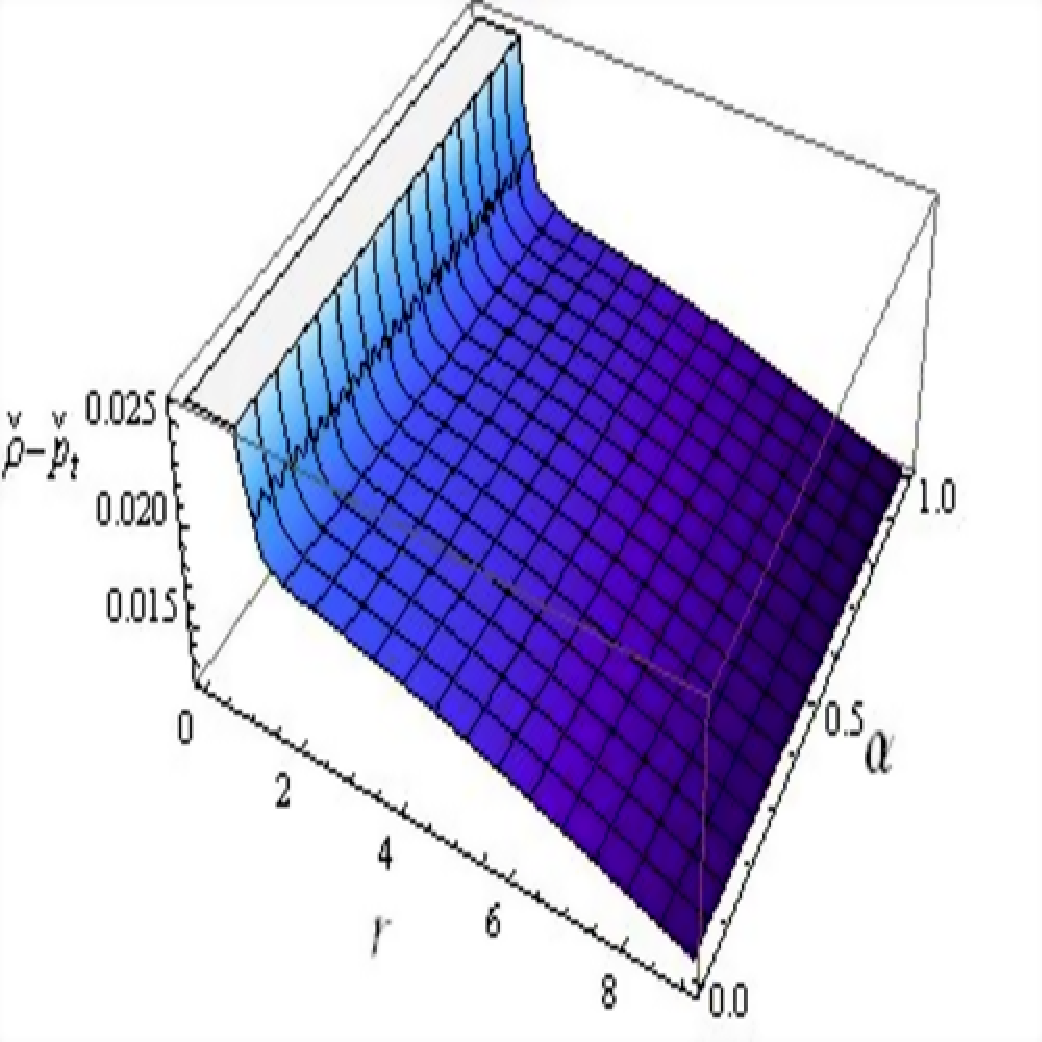,width=0.4\linewidth}
\epsfig{file=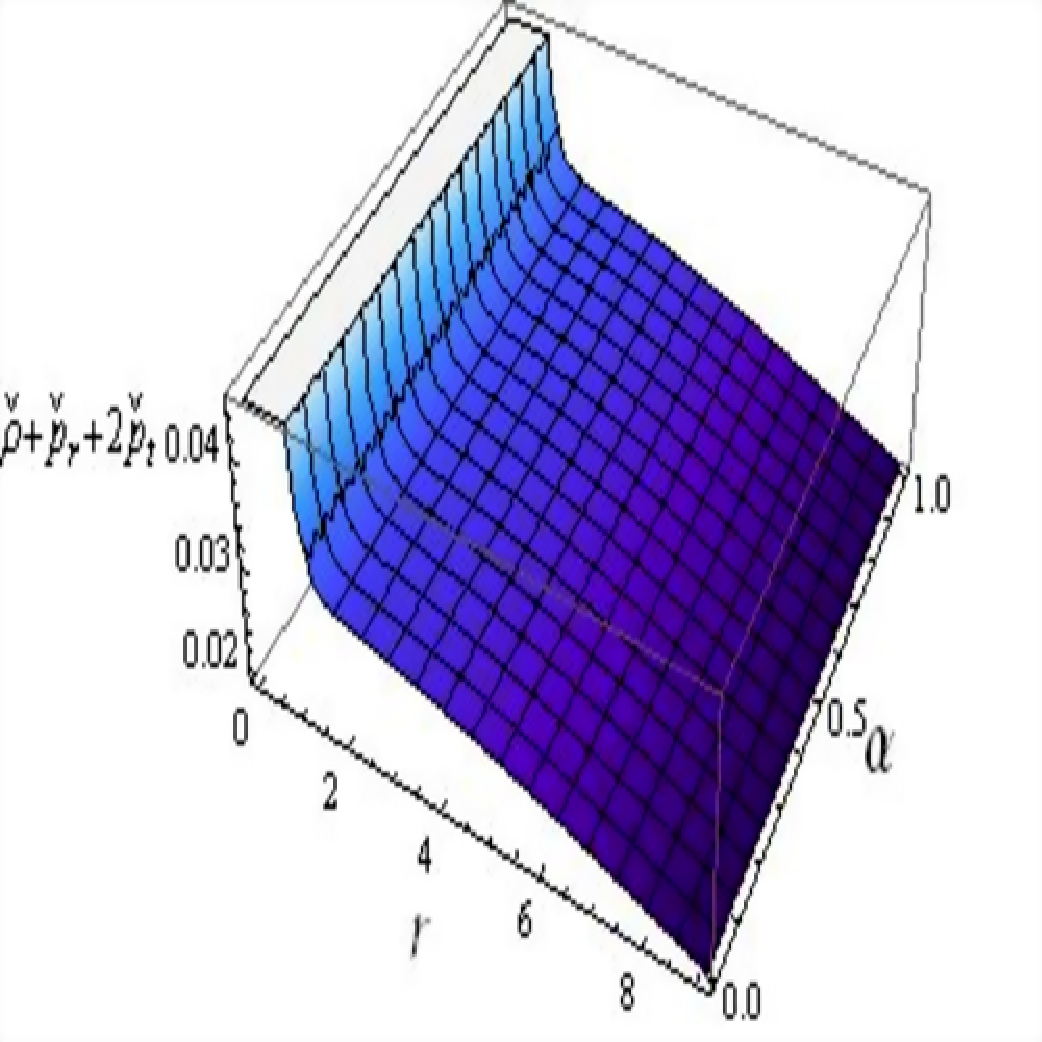,width=0.4\linewidth} \caption{Analysis of energy
constraints for the solution II.}
\end{figure}

The mass of static spherically symmetric distribution is computed by
\begin{equation}\label{58}
m=4\pi\int^{\textsf{R}}_{0}\check{\rho}r^2dr.
\end{equation}
The numerical approach is utilized in Eq.\eqref{58} with the initial
condition $m(0)=0$ to calculate the mass of anisotropic celestial
object. One of the substantial features of an astrophysical object
is its compactness $(\zeta)$, which is defined as the ratio between
mass and radius of the considered star. The upper limit of
compactness is calculated by Buchdahl \cite{22b} by matching the
inner geometry with the outer Schwarzschild vacuum regime through
junction conditions. This limit is found to be less than
$\frac{4}{9}$ for stable stellar configurations. The electromagnetic
radiations are produced by the celestial objects whose wavelength is
enlarged due to strong gravitational pull, thus this increment in
wavelength is analyzed by redshift parameter
$(Z(r)=\frac{1}{\sqrt{1-2\zeta}}-1)$. Buchdahl found this factor as
$Z(r)<2$ for isotropic source but 5.211 for anisotropic matter
\cite{22c}.
\begin{figure}\center
\epsfig{file=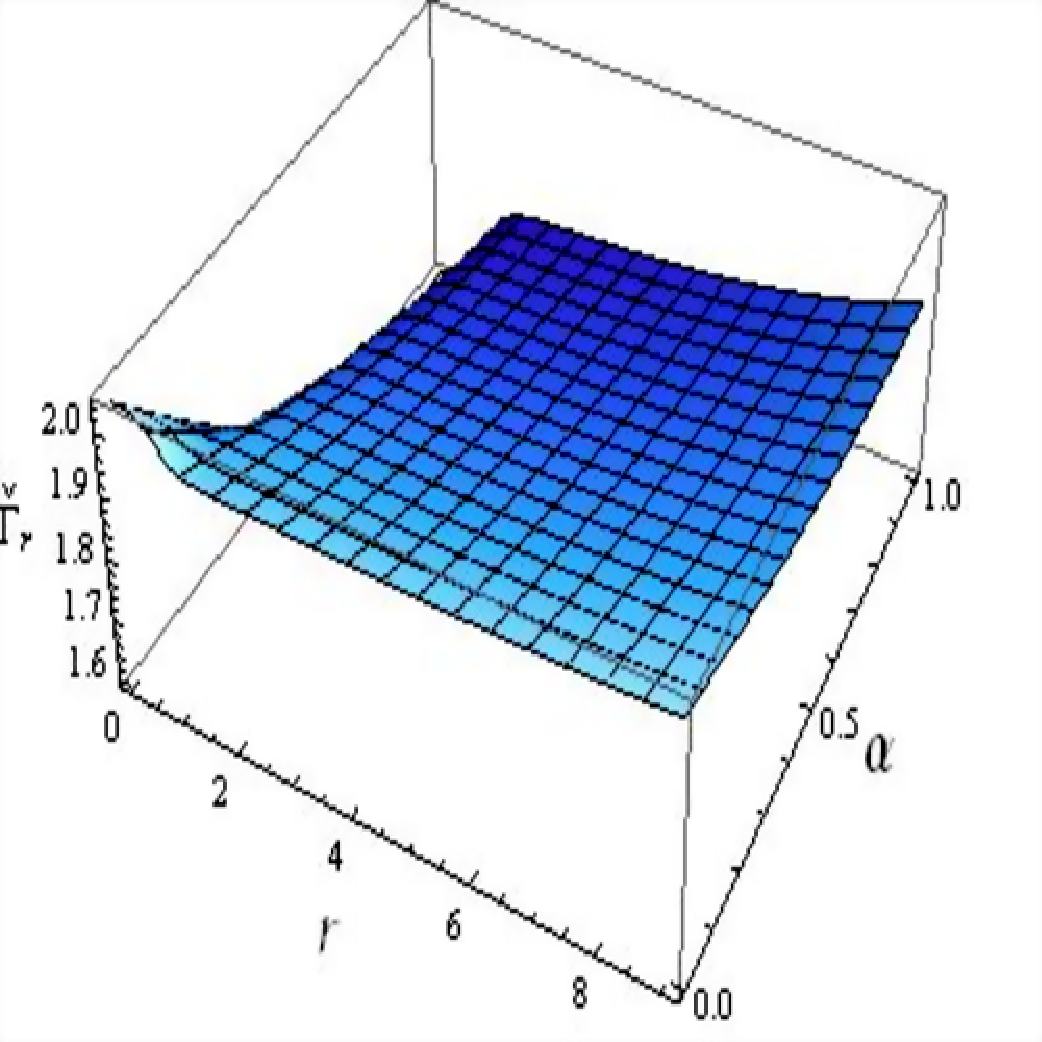,width=0.4\linewidth}\epsfig{file=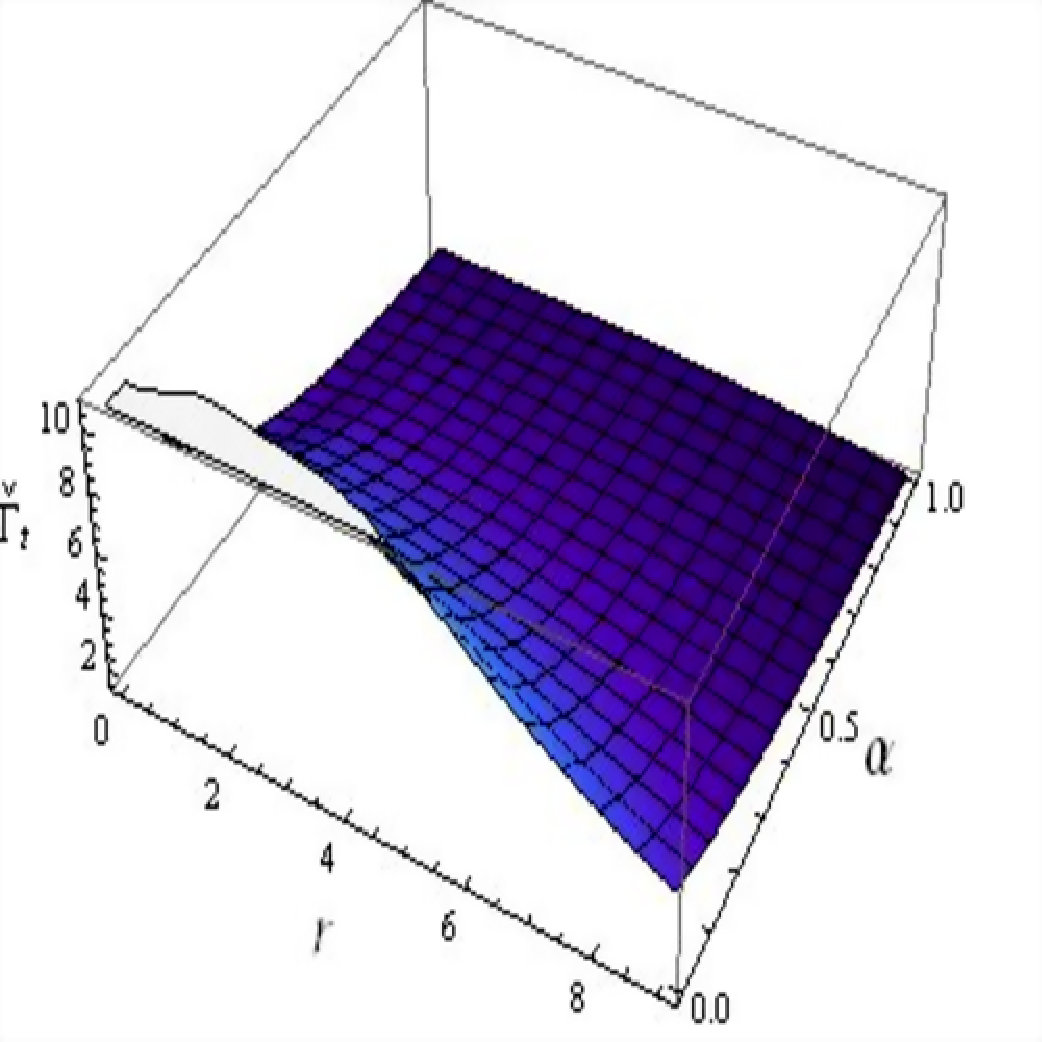,width=0.4\linewidth}
\epsfig{file=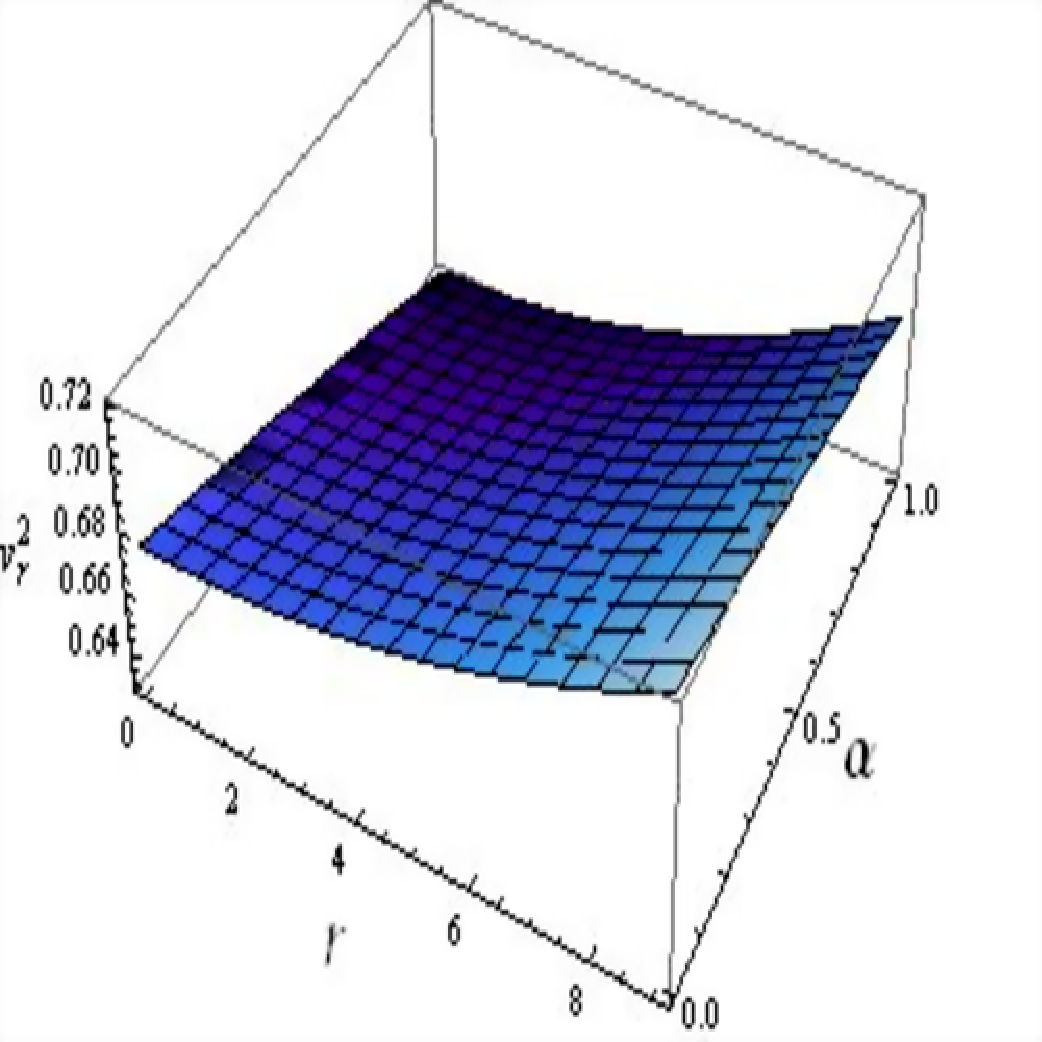,width=0.4\linewidth}\epsfig{file=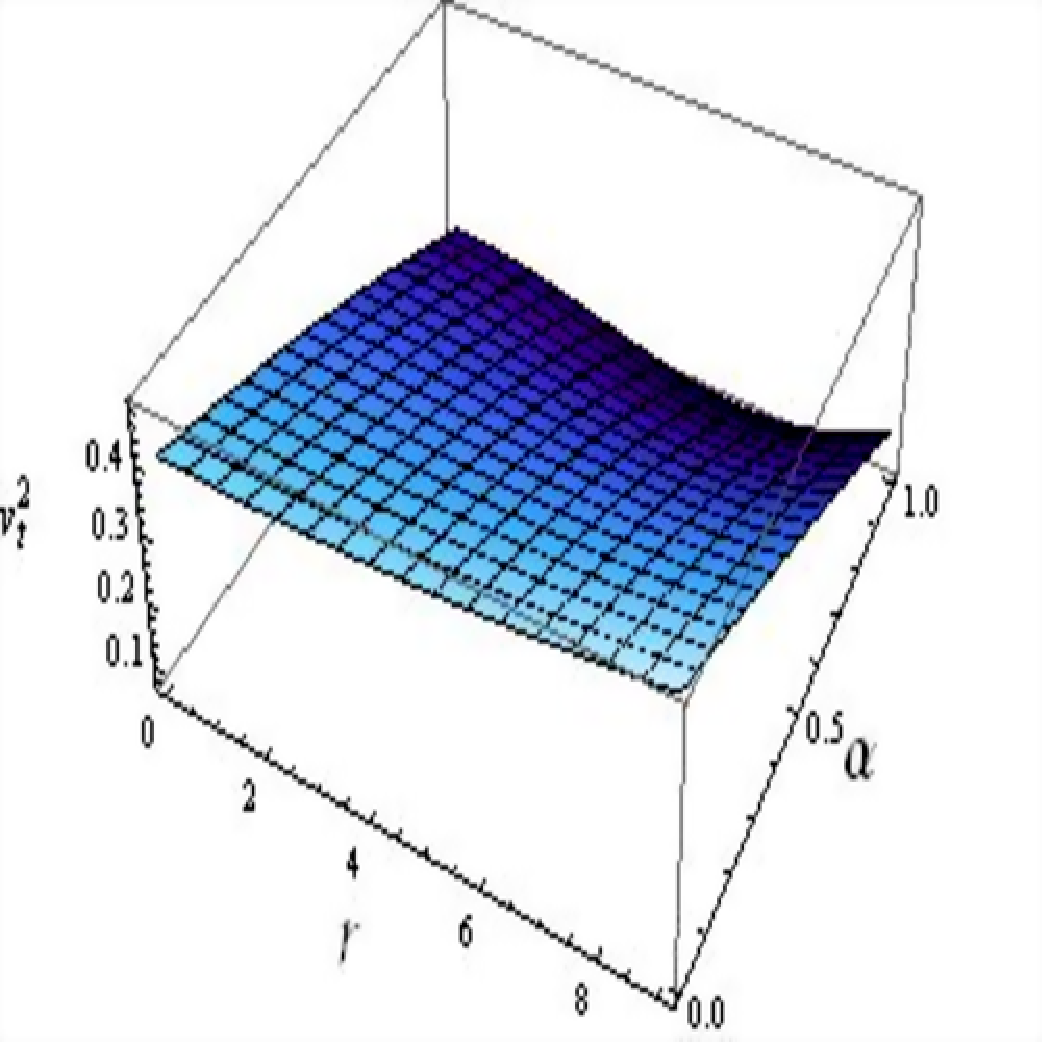,width=0.4\linewidth}
\epsfig{file=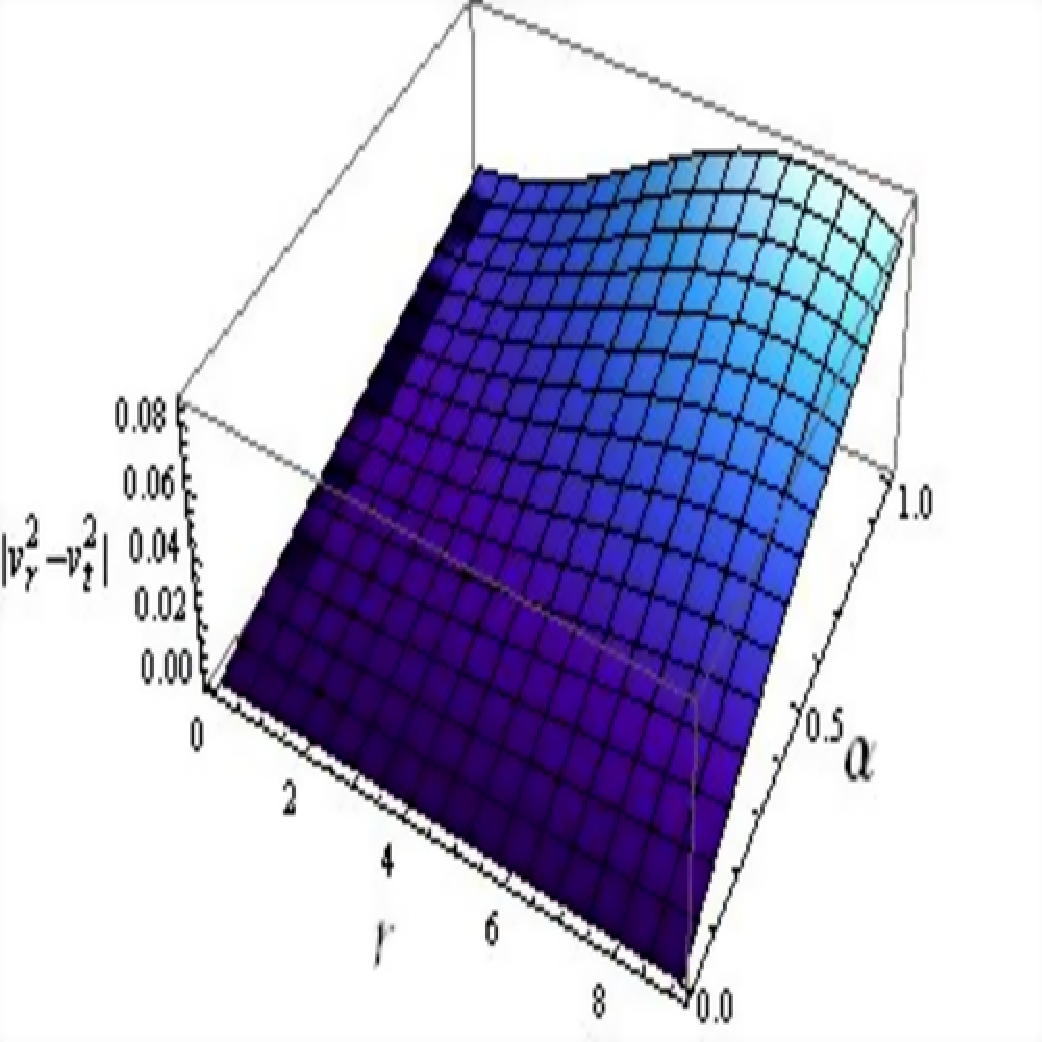,width=0.4\linewidth} \caption{Analysis of
adiabatic index, causality condition and Herrera cracking approach
versus $r$ and $\alpha$ for the solution II.}
\end{figure}
\begin{figure}\center
\epsfig{file=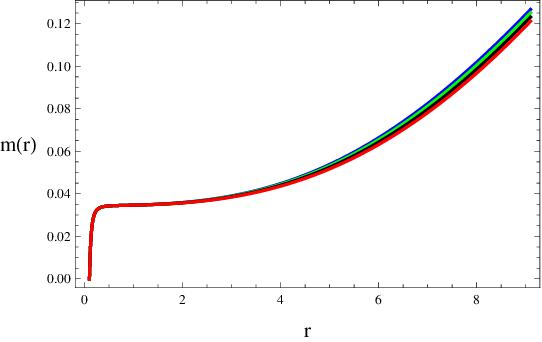,width=0.4\linewidth}\epsfig{file=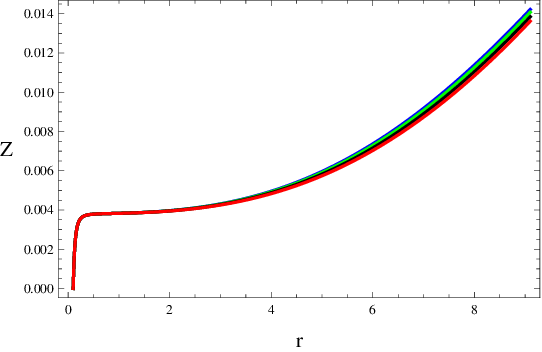,width=0.4\linewidth}
\epsfig{file=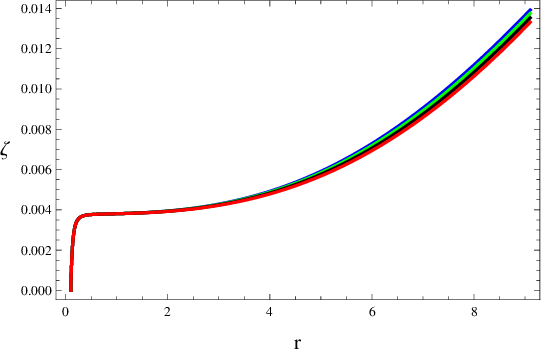,width=0.4\linewidth} \caption{Behavior of mass,
redshift and compactness versus $r$ corresponding to $\alpha=0.01$
(Blue), 0.25 (Green), 0.55 (Black) and 0.85 (Red) for solution I.}
\end{figure}
\begin{figure}\center
\epsfig{file=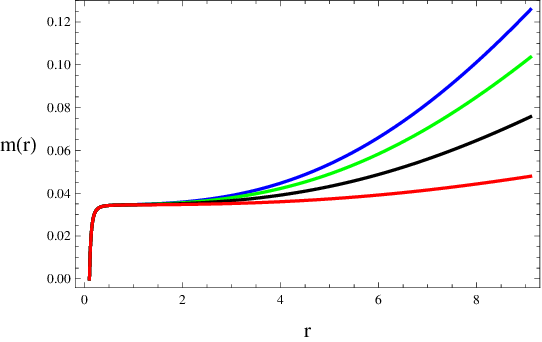,width=0.4\linewidth}\epsfig{file=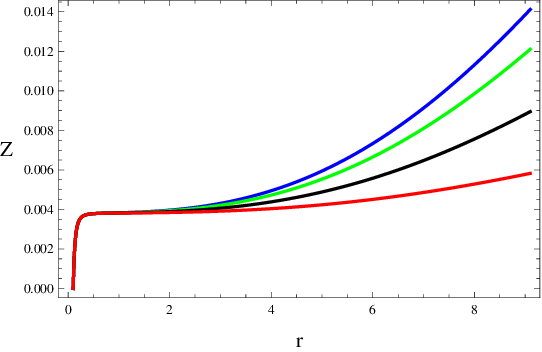,width=0.4\linewidth}
\epsfig{file=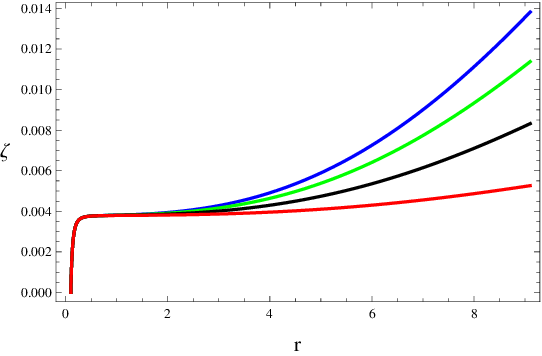,width=0.4\linewidth} \caption{Behavior of mass,
redshift and compactness versus $r$ corresponding to $\alpha=0.01$
(Blue), 0.25 (Green), 0.55 (Black) and 0.85 (Red) for solution II.}
\end{figure}

In order to investigate the mass, compactness and redshift factor
for solutions I and II, we select four values of the decoupling
parameter, i.e., $\alpha=0.01,0.25,0.55,0.85$. Figure \textbf{7}
represents that mass, compactness and redshift parameters slightly
decrease for larger values of $\alpha$ (solution I). For the
solution II, one can notice a significant decline in mass,
compactness as well as redshift factor for higher values of the
decoupling parameter (Figure \textbf{8}). The components of the
equation of state parameter of anisotropic distribution are
evaluated as
\begin{equation}\label{58a}
\check{w_t}=\frac{\check{p_t}}{\check{\rho}}, \quad
\check{w}_{r}=\frac{\check{p_r}}{\check{\rho}}.
\end{equation}
To examine the nature of matter distribution, the components of the
equation of state parameter should be observed from 0 to 1
\cite{23}. Figure \textbf{9} indicates that both these parameters,
for solutions I and II, satisfy the required limit.
\begin{figure}\center
\epsfig{file=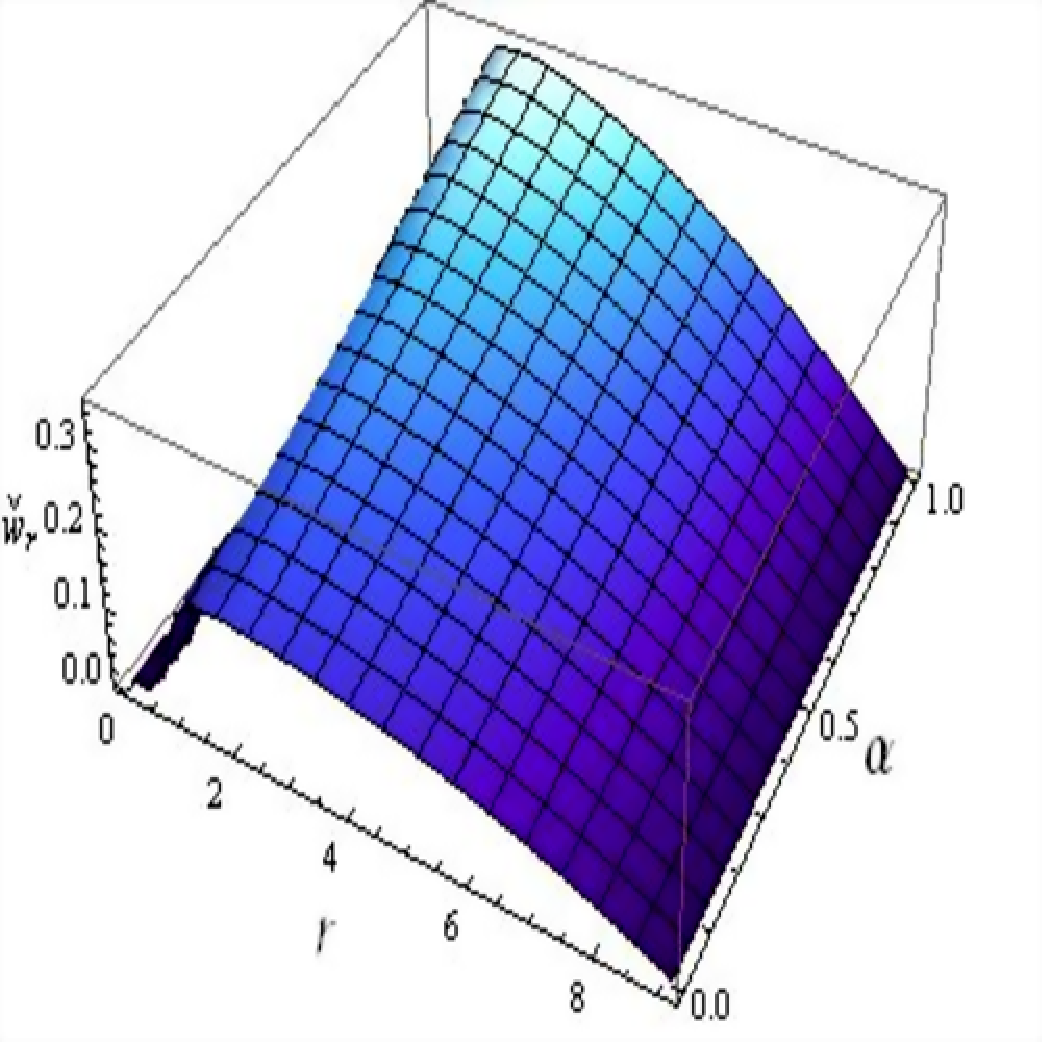,width=0.4\linewidth}\epsfig{file=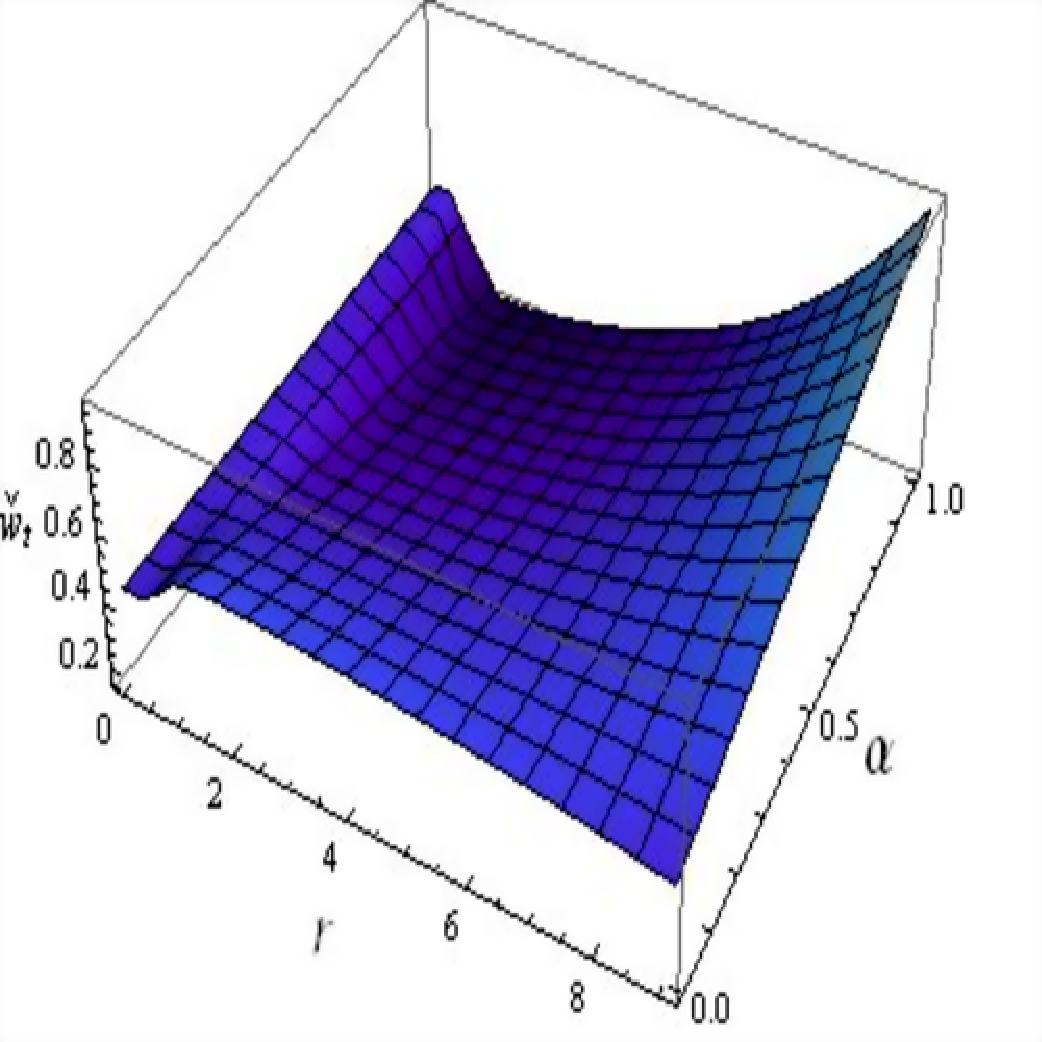,width=0.4\linewidth}
\epsfig{file=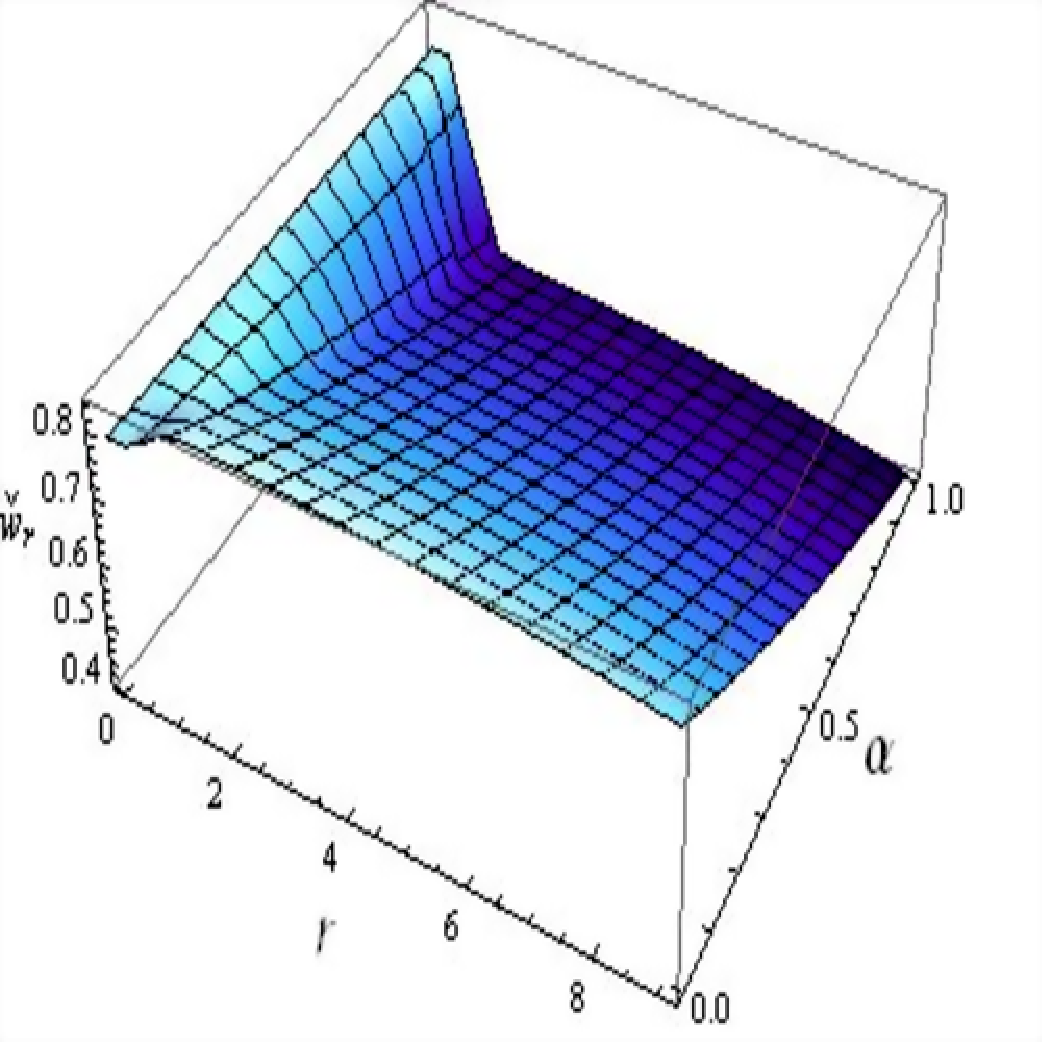,width=0.4\linewidth}\epsfig{file=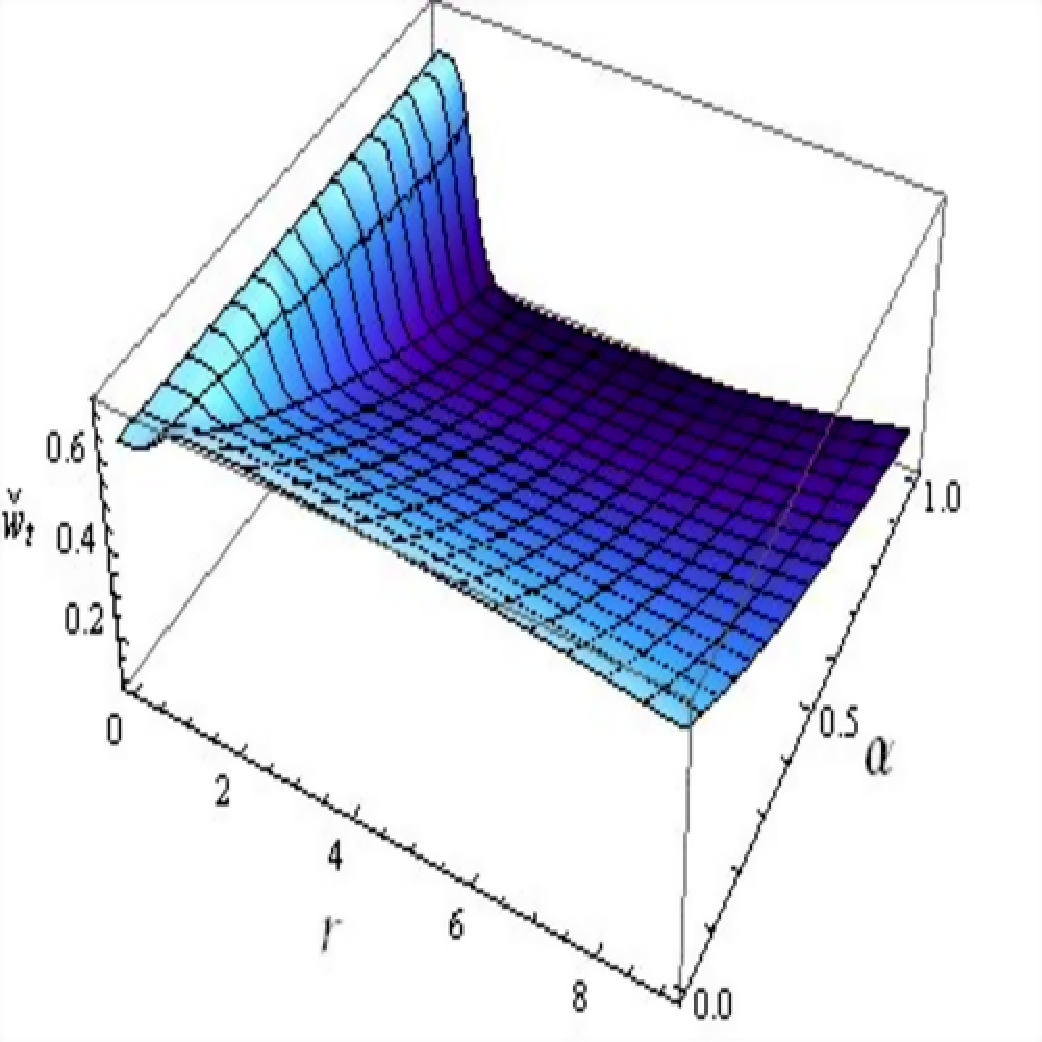,width=0.4\linewidth}
\caption{Analysis of components of the equation of state parameters
versus $r$ and $\alpha$ for solution I and II, respectively.}
\end{figure}

\section{Conclusions}

In the present work, we have developed two anisotropic static
spherical solutions for the model $f(G,T)=\chi G^2+\psi T$ using EGD
scheme. We have induced an extra sector along with the isotropic
configuration to generate anisotropy in the system. The field
equations have been separated into two independent sets by deforming
the radial as well as temporal metric functions, thereby portraying
perfect and anisotropic systems. For the isotropic set, we have
assumed the Krori-Barua ansatz in which the involved unknown
parameters are obtained using matching conditions. The second set
\eqref{21}-\eqref{23} has five unknowns, so we require two more
constraints, the equation of state,
$\delta^{0}_{0}=a\delta^{1}_{1}+b\delta^{2}_{2}$ and pressure-like
or density-like constraint to calculate the anisotropic solutions.
Finally, we have examined the viability and stability of the
resulting solutions through graphs.

We have utilized energy conditions \eqref{59} and three stability
criteria to figure out the viability and stability of the resulting
anisotropic solutions. The feasibility of both solutions has been
confirmed since they meet the limits of energy constraints. Further,
both solutions satisfy all three stability criteria (Herrera
cracking approach, causality condition and adiabatic index), hence
they are stable. The equation of state parameters are also found
consistent for both solutions. The mass, redshift and compactness
parameters are also inspected for $\alpha=0.01,0.25,0.55$ and
$0.85$. The first solution shows little decrement for larger
$\alpha$, whereas a consequential change has been noticed in the
second solution for higher values of $\alpha$.

It is interesting to mention here that two anisotropic solutions
have been constructed in GR \cite{15} which were found to be
unstable in comparison to the present work. In GR, Zubair and Azmat
\cite{20} achieved stable configuration only by causality condition,
while we have developed solutions that are stable in view of all
criteria. Maurya et al. \cite{27} transformed the isotropic source
to the anisotropic distribution in $f(Q)$ gravity (where $Q$ is
non-metricity scalar) and found that stability of the structure is
attained only by keeping $\alpha$ (decoupling parameter) less than
0.18. However, interestingly, it can be seen that our developed
models remain stable throughout the whole domain of $\alpha$. In
$f(R,T)$ gravity, stable anisotropic decoupled solutions have been
generated \cite{26}. Similarly, some viable and stable results have
been found in $f(G)$ gravity \cite{21e}. We have also found
compatible results here. Finally, we would like to mention that all
the results reduce to GR for $\chi=\psi=0$ in the model \eqref{60}.

\section*{Appendix A}
\renewcommand{\theequation}{A\arabic{equation}}
\setcounter{equation}{0} The extra curvature terms in $f(G,T)$ are
given as
\begin{eqnarray}\nonumber
T^{0\textsf{(Cor)}}_{0}&=&\frac{1}{8\pi}\bigg[-\frac{1}{2}
G^2+\bigg(\frac{4 e^{-2 \vartheta} \varphi ''}{r^2}-\frac{4
e^{-\vartheta } \varphi ''}{r^2}-\frac{2 e^{-\vartheta}
\varphi'^2}{r^2}+\frac{2 e^{-\vartheta } \varphi ' \vartheta
'}{r^2}\\\nonumber &+&\frac{2 e^{-2 \vartheta } \varphi
'^2}{r^2}-\frac{6 e^{-2 \vartheta } \varphi ' \vartheta
'}{r^2}\bigg)G+\bigg (\frac{12 e^{-2 \vartheta } \vartheta
'}{r^2}-\frac{4 e^{-\vartheta }\vartheta '}{r^2}\bigg)G'\\\label{61}
&&\bigg (-\frac{8 e^{-2 \vartheta}}{r^2}+\frac{8 e^{-\vartheta }
}{r^2}\bigg)G''\bigg],
\end{eqnarray}
\begin{eqnarray}\nonumber
T^{1\textsf{(Cor)}}_{1}&=&\frac{1}{8 \pi} \bigg[\frac{1}{2} G^2+
\bigg(-\frac{4 e^{-2 \vartheta } \varphi ''}{r^2}+\frac{4
e^{-\vartheta } \varphi ''}{r^2}+\frac{6 e^{-2 \vartheta } \varphi '
\vartheta '}{r^2}-\frac{2 e^{-\vartheta } \varphi ' \vartheta
'}{r^2}\\\nonumber &-&\frac{2 e^{-2 \vartheta } \varphi
'^2}{r^2}+\frac{2 e^{-\vartheta } \varphi '^2}{r^2}\bigg)G+
\bigg(\frac{12 e^{-2\vartheta} \varphi '}{r^2}-\frac{4 e^{-\vartheta
} \varphi '}{r^2}\bigg)G'\bigg],\\\label{62}
\end{eqnarray}
\begin{eqnarray}\nonumber
T^{2\textsf{(Cor)}}_{2}&=&\frac{1}{8 \pi}\bigg[\frac{1}{2} G^2+
\bigg(-\frac{4 e^{-2\vartheta} \varphi ''}{r^2}+\frac{4
e^{-\vartheta } \varphi ''}{r^2}+\frac{2 e^{-\vartheta } \varphi
'^2}{r^2}-\frac{2 e^{-2 \vartheta } \varphi '^2}{r^2}\\\nonumber
&-&\frac{2 e^{-\vartheta } \varphi ' \vartheta '}{r^2}+\frac{6 e^{-2
\vartheta } \varphi ' \vartheta '}{r^2}\bigg)G+ \bigg(-\frac{6 e^{-2
\vartheta } \varphi ' \vartheta '}{r}+\frac{4 e^{-2 \vartheta }
\varphi ''}{r}\\\label{63} &+&\frac{2 e^{-2 \vartheta } \varphi
'^2}{r}\bigg)G'+ \frac{4 e^{-2 \vartheta } \varphi '}{r}G''\bigg].
\end{eqnarray}
The Gauss-Bonnet term as well as its higher derivatives turn out to
be
\begin{eqnarray}\label{63a}
G&=&\frac{1}{r^2}\bigg[2 e^{-2 \vartheta } \left(\left(e^{\vartheta
}-3\right) \vartheta '\varphi ' -\left(2 \vartheta ''+\varphi
'^2\right)\left(e^{\vartheta }-1\right) \right)\bigg],\\\nonumber
G'&=&\frac{-1}{r^3}\bigg[2e^{-2 \vartheta } \big( -\left(\vartheta
''\left(e^{\vartheta }-3\right) -2\varphi '' \left(e^{\vartheta
}-1\right) \right)r \varphi '+r \varphi '\left(e^{\vartheta
}-6\right) \vartheta '^2\big)\\\nonumber &+&\vartheta' \left(r
\left(-\left(e^{\vartheta }-2\right)\right) \varphi '^2+2\varphi 'r
\left(e^{\vartheta }-3\right) -\left(3 e^{\vartheta }-7\right) r
\varphi ''\right)-2\varphi '^2 \big(e^{\vartheta} -1\big)
\\\label{64}&-&2 \left(2 \varphi ''-r \varphi
^{(3)}\right)\left(e^{\vartheta}-1\right) \bigg)\bigg],
\end{eqnarray}
\begin{eqnarray}\nonumber
G''&=&\frac{1}{r^4}\bigg[ 2e^{-2 \vartheta } \bigg(6-6 e^{\vartheta
}+\varphi '^2 \big(r^2 \big(e^{\vartheta }-2\big) \vartheta
''\big)-2 \bigg(\varphi '' \big(6 \big(e^{\vartheta }-1\big)-r^2
\big(2 e^{\vartheta }\\\nonumber&-&5\big) \vartheta ''\big)+\varphi
''^2r^2 \big(e^{\vartheta }-1\big) +r \big(-4 \varphi ^{(3)}+r
\varphi ^{(4)}\big) \big(e^{\vartheta }-1\big)\bigg)+r^2
\big(e^{\vartheta }-12\big)\\\nonumber&\times& \varphi ' \vartheta
'^3+\vartheta ' \bigg(\varphi ' \big(-3 r^2 \big(e^{\vartheta
}-6\big) \vartheta ''+4 r^2 \big(e^{\vartheta }-2\big) \varphi ''+6
\big(e^{\vartheta }-3\big)\big)-4\\\nonumber&\times& r\varphi '^2
\big(e^{\vartheta }-2\big) +r \big( \big(5 e^{\vartheta }-11\big)r
\varphi ^{(3)}-4 \varphi ''\big(3 e^{\vartheta }-7\big)\big)\bigg)-r
\vartheta '^2 \bigg(4 r \big(e^{\vartheta }\\\nonumber&-5&\big)
\varphi ''-4 \big(e^{\vartheta }-6\big) \varphi '\bigg)+r
\big(e^{\vartheta }-4\big) \varphi '^2+ \varphi ' r \bigg(r
\big(\big(e^{\vartheta }-3\big) \vartheta ^{(3)}-2 \varphi
^{(3)}\\\label{65}&\times& \big(e^{\vartheta }-1\big)\big)-4
\big(e^{\vartheta }-3\big) \vartheta ''+8 \big(e^{\vartheta }-1\big)
\varphi ''\bigg)\bigg)\bigg].
\end{eqnarray}
\section*{Appendix B}
\renewcommand{\theequation}{B\arabic{equation}}
\setcounter{equation}{0} The radial component of adiabatic index
corresponding to solutions I and II are
\begin{align}\nonumber
\check{\Gamma_{r}}&=\{8 \pi  (\psi +4 \pi ) r^2 \big(2
(L+X)-T^{0\textsf{(Cor)}}_{0} e^{r^2 X}-T^{1\textsf{(Cor)}}_{1}
e^{r^2 X}\big) \big(\alpha  \psi ^2+12 \pi  \alpha  \psi
\\\nonumber&+32 \pi ^2 \alpha +4 \pi  \psi +2 \alpha \psi ^2 L r^4 X+24
\pi  \alpha \psi L r^4 X+64 \pi ^2 \alpha L r^4 X+12 \pi  \psi  L
r^4 X\\\nonumber&+64 \pi ^2 L r^4 X+4 \pi \psi r^4 X^2+\alpha \psi
^2 r^2 X-\alpha  \psi ^2 e^{r^2 X}+12 \pi \alpha \psi r^2 X-12 \pi
\alpha \psi e^{r^2 X}\\\nonumber&+32 \pi ^2 \alpha  r^2 X-32 \pi ^2
\alpha e^{r^2 X}+4 \pi \psi  r^2 X-4 \pi \psi e^{r^2 X}+32 \pi ^2
r^2 X-32 \pi ^2 e^{r^2 X}\\\nonumber&+\pi \psi r^3 e^{r^2 X}
{T^{0\textsf{(Cor)}}_{0}}^{'}+\pi (3 \psi +16 \pi ) r^3 e^{r^2 X}
{T^{1\textsf{(Cor)}}_{1}}^{'}+32 \pi ^2\big)\}\{\big(\alpha \psi
^2+12 \pi \alpha \psi\\\nonumber& +32 \pi ^2 \alpha +4 \pi  \psi +2
\alpha \psi ^2 L r^2+24 \pi \alpha \psi  L r^2+64 \pi ^2 \alpha  L
r^2+12 \pi \psi  L r^2+64 \pi ^2 L r^2\\\nonumber&-2 \pi \psi r^2
T^{0\textsf{(Cor)}}_{0} e^{r^2 X}-2 \pi (3 \psi +16 \pi ) r^2
T^{1\textsf{(Cor)}}_{1} e^{r^2 X}-\alpha \psi ^2 e^{r^2 X}-12 \pi
\alpha \psi  e^{r^2 X}\\\nonumber&-32 \pi ^2 \alpha e^{r^2 X}+4 \pi
\psi r^2 X-4 \pi  \psi  e^{r^2 X}-32 \pi ^2 e^{r^2 X}+32 \pi ^2\big)
\big(-\alpha \psi ^2-12 \pi \alpha  \psi \\\nonumber&-32 \pi ^2
\alpha -4 \pi \psi -2 \alpha \psi ^2 L r^4 X-24 \pi \alpha \psi L
r^4 X-64 \pi ^2 \alpha L r^4 X+4 \pi  \psi L r^4 X\\\nonumber&+12
\pi \psi r^4 X^2+64 \pi ^2 r^4 X^2-\alpha \psi ^2 r^2 X+\alpha \psi
^2 e^{r^2 X}-12 \pi \alpha \psi  r^2 X+12 \pi \alpha \psi e^{r^2
X}\\\nonumber&-32 \pi ^2 \alpha r^2 X+32 \pi ^2 \alpha e^{r^2 X}-4
\pi \psi r^2 X+4 \pi \psi e^{r^2 X}-32 \pi ^2 r^2 X+32 \pi ^2 e^{r^2
X}\\\label{67}&+\pi (3 \psi +16 \pi ) r^3 e^{r^2 X}
{T^{0\textsf{(Cor)}}_{0}}^{'}+\pi \psi r^3 e^{r^2 X}
{T^{1\textsf{(Cor)}}_{1}}^{'}-32 \pi ^2\big)\}^{-1},
\end{align}
\begin{align}\nonumber
\check{\Gamma_{r}}&=\{8 \pi  (\psi +4 \pi ) r^2 \big(2
(L+X)-T^{0\textsf{(Cor)}}_{0} e^{r^2 X}-T^{1\textsf{(Cor)}}_{1}
e^{r^2 X}\big) \big(-\alpha \psi ^2-12 \pi  \alpha \psi
\\\nonumber&-32 \pi ^2 \alpha +4 \pi  \psi +12 \pi \psi L r^4 X+64
\pi ^2 L r^4 X+2 \alpha \psi ^2 r^4 X^2+24 \pi \alpha \psi  r^4
X^2\\\nonumber&+64 \pi ^2 \alpha r^4 X^2+4 \pi \psi r^4 X^2-\alpha
\psi ^2 r^2 X+\alpha \psi ^2 e^{r^2 X}+32 \pi ^2 \alpha e^{r^2 X}+4
\pi  \psi r^2 X\\\nonumber&-12 \pi \alpha \psi r^2 X+12 \pi  \alpha
\psi  e^{r^2 X}-32 \pi ^2 \alpha  r^2 X-4 \pi \psi e^{r^2 X}+32 \pi
^2 r^2 X-32 \pi ^2 e^{r^2 X}\\\nonumber&+\pi \psi r^3 e^{r^2 X}
{T^{0\textsf{(Cor)}}_{0}}^{'}+\pi (3 \psi +16 \pi ) r^3 e^{r^2 X}
{T^{1\textsf{(Cor)}}_{1}}^{'}+32 \pi ^2\big)\}\{\big(-\alpha \psi
^2-12 \pi \alpha \psi\\\nonumber& -32 \pi ^2 \alpha +4 \pi  \psi +12
\pi \psi  L r^2+64 \pi ^2 L r^2-2 \pi \psi r^2
T^{0\textsf{(Cor)}}_{0} e^{r^2 X}-2 \pi  (3 \psi +16 \pi )
r^2\\\nonumber&\times T^{1\textsf{(Cor)}}_{1} e^{r^2 X}+2 \alpha
\psi ^2 r^2 X+\alpha \psi ^2 e^{r^2 X}+24 \pi \alpha \psi r^2 X+12
\pi \alpha \psi e^{r^2 X}+64 \pi ^2 \alpha r^2 X\\\nonumber&+32 \pi
^2 \alpha e^{r^2 X}+4 \pi \psi r^2 X-4 \pi \psi e^{r^2 X}-32 \pi ^2
e^{r^2 X}+32 \pi ^2\big) \big(\alpha \psi ^2+12 \pi \alpha \psi +32
\pi ^2 \alpha \\\nonumber&-4 \pi \psi +4 \pi \psi  L r^4 X-2 \alpha
\psi ^2 r^4 X^2-24 \pi \alpha \psi r^4 X^2-64 \pi ^2 \alpha r^4
X^2+12 \pi \psi r^4 X^2\\\nonumber&+64 \pi ^2 r^4 X^2+\alpha \psi ^2
r^2 X-\alpha \psi ^2 e^{r^2 X}+12 \pi \alpha \psi r^2 X-12 \pi
\alpha \psi e^{r^2 X}+32 \pi ^2 \alpha r^2 X\\\nonumber&-32 \pi ^2
\alpha e^{r^2 X}-4 \pi  \psi r^2 X+4 \pi \psi e^{r^2 X}-32 \pi ^2
r^2 X+32 \pi ^2 e^{r^2 X}+\pi (3 \psi +16 \pi ) r^3
\\\label{68}&\times e^{r^2 X} {T^{0\textsf{(Cor)}}_{0}}^{'}+\pi \psi r^3
e^{r^2 X} {T^{1\textsf{(Cor)}}_{1}}^{'}-32 \pi ^2\big)\}^{-1}.
\end{align}
The expressions of radial velocity in the case of first and second
solution are given as
\begin{align}\nonumber
\nu^{2}_{r}&=\{\alpha  \psi ^2+12 \pi  \alpha  \psi +32 \pi ^2
\alpha +4 \pi  \psi +2 \alpha  \psi ^2 L r^4 X+24 \pi \alpha \psi L
r^4 X+64 \pi ^2 \alpha  L r^4 X\\\nonumber&+12 \pi  \psi  L r^4 X+64
\pi ^2 L r^4 X+4 \pi  \psi  r^4 X^2+\alpha  \psi ^2 r^2 X-\alpha
\psi ^2 e^{r^2 X}+12 \pi  \alpha \psi  r^2 X\\\nonumber&-12 \pi
\alpha \psi e^{r^2 X}+32 \pi ^2 \alpha  r^2 X-32 \pi ^2 \alpha
e^{r^2 X}+4 \pi \psi r^2 X-4 \pi \psi e^{r^2 X}+32 \pi ^2 r^2
X\\\nonumber&-32 \pi ^2 e^{r^2 X}+\pi \psi r^3 e^{r^2 X}
{T^{0\textsf{(Cor)}}_{0}}^{'}+\pi (3 \psi +16 \pi ) r^3 e^{r^2 X}
{T^{1\textsf{(Cor)}}_{1}}^{'}+32 \pi ^2\}\{-\alpha \psi
^2\\\nonumber&-12 \pi \alpha \psi -32 \pi ^2 \alpha -4 \pi  \psi -2
\alpha \psi ^2 L r^4 X-24 \pi \alpha \psi L r^4 X-64 \pi ^2 \alpha L
r^4 X\\\nonumber&+4 \pi \psi L r^4 X+12 \pi \psi r^4 X^2+64 \pi ^2
r^4 X^2-\alpha \psi ^2 r^2 X+\alpha \psi ^2 e^{r^2 X}-12 \pi \alpha
\psi r^2 X\\\nonumber&+12 \pi \alpha \psi e^{r^2 X}-32 \pi ^2 \alpha
r^2 X+32 \pi ^2 \alpha e^{r^2 X}-4 \pi \psi r^2 X+4 \pi \psi e^{r^2
X}-32 \pi ^2 r^2 X\\\label{69}&+32 \pi ^2 e^{r^2 X}+\pi (3 \psi +16
\pi ) r^3 e^{r^2 X} {T^{0\textsf{(Cor)}}_{0}}^{'}+\pi \psi r^3
e^{r^2 X} {T^{1\textsf{(Cor)}}_{1}}^{'}-32 \pi ^2\}^{-1},
\end{align}
\begin{align}\nonumber
\nu^{2}_{r}&=\{-\alpha  \psi ^2-12 \pi  \alpha  \psi -32 \pi ^2
\alpha +4 \pi  \psi +12 \pi  \psi  L r^4 X+64 \pi ^2 L r^4 X+2
\alpha  \psi ^2 r^4 X^2\\\nonumber&+24 \pi  \alpha  \psi  r^4 X^2+64
\pi ^2 \alpha  r^4 X^2+4 \pi  \psi  r^4 X^2-\alpha \psi ^2 r^2
X+\alpha \psi ^2 e^{r^2 X}-12 \pi  \alpha \psi r^2 X\\\nonumber&+12
\pi \alpha \psi e^{r^2 X}-32 \pi ^2 \alpha  r^2 X+32 \pi ^2 \alpha
e^{r^2 X}+4 \pi \psi  r^2 X-4 \pi  \psi  e^{r^2 X}+32 \pi ^2 r^2
X\\\nonumber&-32 \pi ^2 e^{r^2 X}+\pi  \psi  r^3 e^{r^2 X}
{T^{0\textsf{(Cor)}}_{0}}^{'}+\pi (3 \psi +16 \pi ) r^3 e^{r^2 X}
{T^{1\textsf{(Cor)}}_{1}}^{'}+32 \pi ^2\}\{\alpha \psi
^2\\\nonumber&+12 \pi \alpha \psi  -4 \pi \psi +4 \pi \psi  L r^4
X-2 \alpha \psi ^2 r^4 X^2-24 \pi \alpha \psi r^4 X^2-64 \pi ^2
\alpha r^4 X^2\\\nonumber&+12 \pi \psi r^4 X^2+32 \pi ^2 \alpha+64
\pi ^2 r^4 X^2+\alpha \psi ^2 r^2 X-\alpha \psi ^2 e^{r^2 X}+12 \pi
\alpha \psi r^2 X\\\nonumber&-12 \pi \alpha \psi e^{r^2 X}+32 \pi ^2
\alpha  r^2 X-32 \pi ^2 \alpha e^{r^2 X}-4 \pi \psi r^2 X+4 \pi \psi
e^{r^2 X}-32 \pi ^2 r^2 X\\\label{70}&+32 \pi ^2 e^{r^2 X}+\pi (3
\psi +16 \pi ) r^3 e^{r^2 X} {T^{0\textsf{(Cor)}}_{0}}^{'}+\pi \psi
r^3 e^{r^2 X} {T^{1\textsf{(Cor)}}_{1}}^{'}-32 \pi ^2\}^{-1}.
\end{align}

\end{document}